\documentclass{IEEEtran}

\usepackage{subcaption}
\usepackage{dsfont}
\usepackage{amsmath,hhline}
\usepackage{color}
\usepackage{multirow}
\usepackage{url,cite}
\usepackage{lastpage,algorithm,algorithmic,graphicx}
\usepackage{epsfig,amssymb}
\usepackage{bbding}

\begin{document}

\title{Learning Based Frequency- and Time-Domain Inter-Cell Interference Coordination in HetNets}
\author{\IEEEauthorblockN{Meryem Simsek$^\ddag$, Mehdi Bennis$^\dag$,  and \.{I}smail G\"uven\c{c}$^\ddag$\\}
\IEEEauthorblockA{$^\ddag$Department of Electrical \& Computer Engineering, Florida International University, USA\\
$^\dag$Centre for Wireless Communications, University of Oulu, Finland\\
Email: meryem.simsek@ieee.org,  bennis@ee.oulu.fi, and iguvenc@fiu.edu}
}

\maketitle
\vspace{-0.9cm}
\begin{abstract}
In this article, we focus on  inter-cell interference coordination  (ICIC) techniques in  heterogeneous network (HetNet) deployments, whereby  macro- and picocells \emph{autonomously} optimize their downlink transmissions, with loose coordination. We model this \emph{strategic} coexistence as a multi-agent system, aiming at joint  interference management and  cell association.  Using tools from Reinforcement Learning (RL),  agents (i.e., macro- and picocells) sense their environment, and self-adapt based on local information so as to maximize their network performance. Specifically, we explore  both time- and frequency domain ICIC scenarios, and propose a \emph{two-level} RL formulation. Here, picocells learn their optimal cell range expansion (CRE) bias and transmit power allocation, as well as appropriate frequency bands for multi-flow transmissions, in which a user equipment (UE) can be simultaneously served by two or more base stations (BSs) from macro- and pico-layers. 
To substantiate our theoretical findings, Long Term Evolution Advanced (LTE-A) based system level simulations are carried out in which our proposed approaches are compared with a number of baseline approaches, such as resource partitioning (RP), static CRE, and single-flow Carrier Aggregation (CA). Our proposed solutions yield substantial gains up to $125\%$ compared to static ICIC approaches in terms of average UE throughput in the time-domain. In the frequency-domain our proposed solutions yield gains up to 240\% in terms of cell-edge UE throughput.
\end{abstract}



%
\IEEEpeerreviewmaketitle
Learning Based Frequency- and Time-Domain Inter-Cell Interference Coordination in HetNets

\section{Introduction}

Driven by the  network densification and increasing number of  smart-phones, tablets and netbooks, mobile operators are compelled to find viable solutions to maximize their network performance in a cost-effective manner. Heterogeneous network (HetNets) deployments combining various cell sizes (femto, pico, relays) and radio access technologies ($3$G/$4$G/Wi-Fi), are expected to become cornerstones for future heterogeneous wireless cellular networks, aiming at substantially higher data rates and  spatial reuse \cite{Ghosh}. HetNets are currently studied within the $3^{\text{rd}}$ Generation Partnership Project ($3$GPP) standardization body, where mechanisms including time- and frequency-domain intercell interference coordination (ICIC) with adaptive resource partitioning, cell range expansion (CRE), and interference coordination/cancellation take central stage \cite{Damjanovic}. In this article, we focus on one of these important aspects, namely self-organizing networks (SON). $3$GPP has defined SON as one of the most important standardization features for mobile operators today for the operation, management,  and maintenance of their radio access networks (RANs) cost-efficiently, without relying on human intervention \cite{Hamalainen}. SON in HetNets is expected to gain more importance as networks are getting denser and becoming more heterogeneous in size, access technology, and backhauls. Endowed with self-configuring, self-optimizing and self-healing capabilities, mobile operators  can optimize their networks in a totally decentralized manner, in which the traffic load is balanced among tiers, significantly reducing their operation and capital expenditures (OPEX/CAPEX), and ultimately satisfying  users' quality-of-service (QoS) requirements.

Based on the self-organizing capabilities of HetNets, we propose solutions to \emph{smartly} offload traffic to open access picocells and thereby achieve cell splitting gains for both time- and frequency-domain ICIC techniques. We focus on the downlink transmission as this has been identified as a more critical intercell interference scenario within HetNets~\cite{Lopez}. Open access picocells are cells that provide access to any user equipment (UE) within their coverage area. As UEs generally connect to the cell that provides the strongest downlink (DL) received signal, DL intercell interference can be reduced. However, if all UEs connect to the macrocell due to their large transmit power, rather than to picocells at shorter distance with lesser number of UEs, the traffic load will be unevenly distributed in the network. As a result, the macrocell will be overloaded whereas picocells will be under-utilized~\cite{Lopez1,Guvenc,Brueck}. As a remedy to this, the concept of CRE was proposed as a cell selection procedure, in which a positive bias is added to the picocell's DL received signal to increase its DL footprint. This bias balances the load among the macro- and picocell tier by forcing mobile users to handover to picocells, even if the picocell's DL received signal is lower. Nevertheless, an aggressive range expansion may cause high interference to picocell UEs (PUEs) located in the picocell expanded regions (ER); this is because ER PUEs do not connect to the cells with the strongest DL received signal, thus suffering from low DL Signal-to-Interference-plus-Noise Ratios (SINRs)~\cite{Madan}. In addition, due to the non-uniform traffic and user distribution, picocells need to self-organize for effectively offloading macrocell's traffic. With this in mind, \emph{intelligent} and flexible cell range expansion techniques across time and frequency must be devised for macro- and picocells, to mitigate excessive DL inter-cell interference suffered by ER PUEs, while at the same time not jeopardizing PUE QoS requirements.

\section{Related Work and Contributions}
In this section, we summarize the concepts of  range expansion and time/frequency domain ICIC  in HetNets and discuss related works from the literature to better present our contributions.
\subsection{Picocell Range Expansion and Inter-Cell Interference Coordination}
In order to benefit from the deployment of heterogeneous and small cell networks, range expansion ICIC techniques have been proposed, in which picocells increase their footprint so as to balance the load among tiers and achieve cell-splitting gains. In what follows, we revisit and summarize both the range expansion and ICIC concepts.

$3$GPP has studied the concept of CRE through handover biasing and resource partitioning among nodes with different levels of transmission powers \cite{3GPP1,3GPP2,3GPP}. The biasing mechanism allows load balancing among tiers, where depending on the bias value, more UEs can be associated to picocells. In this approach, the bias value is an offset  added to the received power of picocells in order to increase its DL coverage area. CRE significantly reduces the DL signal quality of those users in the expanded region (i.e., ER PUEs), because they are connected to cells that do not provide the best DL received signal. These interference problems may significantly degrade the overall network performance, calling for intelligent ICIC schemes to  benefit from range expansion  and improve the performance of ER PUEs. Since ICIC schemes specified in $3$GPP LTE Release $8-9$ do not specifically consider HetNet settings, enhancements of these techniques have been proposed to efficiently mitigate interference in subsequent releases of the LTE standard~\cite{CMCC}. In particular, the ICIC techniques in $3$GPP Release $10-12$, can be grouped into four categories: time-domain, frequency-domain, power based and antenna/spatial-based techniques \cite{3GPP4,3GPP5}.

\subsection{Literature Review}
There is a sizeable body of literature on the use of CRE for traffic load balancing in HetNets; see e.g. \cite{Lopez1,Guvenc,Damjanovic2,Madan,Brueck,Mukherjee,Shirakabe,Merwaday1,Merwaday2,Merwaday3} and the references listed therein. In \cite{Lopez1}, closed-form expressions are derived to calculate CRE bias values for different range expansion strategies. Moreover, a cooperative scheduling scheme is proposed to mitigate interference caused by macrocells onto ER PUEs. To improve DL capacity and users' fairness, the authors propose a new subframe blanking based cell selection procedure in \cite{Guvenc}. Using tools from stochastic geometry, analytical models accounting for base station (BS) and UE locations have been studied to  analyze spectral efficiencies in range expanded picocell networks in \cite{Mukherjee}, which has later been extended to ICIC scenarios in~\cite{Merwaday1,Merwaday2,Merwaday3}.
 In \cite{Shirakabe}, the throughput performance of different CRE values and different ratios of protected resources were carried out based on system level simulations.

In addition to time domain interference coordination approaches, frequency domain interference coordination techniques have also been considered in the literature for interference management and load balancing purposes. In this context, multi-flow carrier aggregation (CA), in which users are served by different layers on different component carriers (CCs), has and remains an open and challenging problem. A related approach to provide an efficient and flexible network performance
improvement is to split the control and user plane (C-and U-plane). This concept was introduced and discussed in \cite{Docomo,Ishii} whereby, the C-plane is provided at low frequency band to maintain good connectivity and mobility.
On the other hand, the U-plane is provided by both the macrocells and the small cells (deployed at higher frequency bands) for data transfer. Since small cells are not configured with cell-specific signals and channels, they
are named \emph{Phantom Cells}~\cite{Ishii}.

\subsection{Contribution}
The main contribution of this article is to propose decentralized solutions for joint power control and cell association in a HetNet scenario, in both time and frequency domain. In the time-domain,  Pico Base Stations (PBSs) \emph{optimally} learn their CRE bias and power allocation, while satisfying their own PUEs' QoS requirements. In turn, the macrocell self-organizes so as to serve its own macro UEs (MUEs), while adhering to the picocell interference constraint. In contrast to the \emph{homogeneous} case where all PBSs use the same bias value, the proposed solution is dynamic and self-organizing in nature, where the RAN autonomously optimizes the CRE bias values of the picocells through a loose coordination with the macrocell tier.  The UE adds these bias values to its measurements, to check whether a measurement report needs to be sent to its serving BS. The PBSs, upon coordination with the MBS, learn the CRE bias values and notify the MBS via the X2 interface. 

In the frequency-domain, we consider: (a) the single-flow CA, where users are served by only one BS at a time, and (b) the multi-flow CA, in which a UE can be simultaneously served by two (or more) BSs from different layers/tiers, but on two different CCs. Our proposed learning based solution is validated using a long term evolution advanced (LTE-A) system level simulator, through a comparison with a number of benchmark solutions such as resource partitioning and static CRE.

It is worth noting that most of the existing ICIC and load balancing techniques are simulated in simplified HetNet scenarios with homogeneity inside the macro layer as well as the pico layer; by considering the same CRE for all picocells in the network. The major difference between our contribution and existing techniques is that we propose a joint optimization approach, in which each picocell individually learns its optimum ICIC strategy. This is achieved by optimizing the picocell’s CRE bias selection and power allocation strategies in coordination with the macrocell. In contrast to existing approaches, our solution is based on Reinforcement learning, which is a widely accepted tool in dynamic wireless networks and allows to investigate how BSs interact over time and attempt to optimize their utility~\cite{Han}.  We propose a reinforcement learning framework, in which not only the picocells but also the macrocell perform load balancing and power control. The challenge of this approach lies in effectively offloading the UEs from the macrocells, while simultaneously maintaining the QoS requirements of PUEs. By enabling coordination between both layers and considering the performance of MUEs and PUEs, the proposed techniques are seen as a promising approach to overcome this challenge.

The remainder of this paper is organized as follows: Section III summarizes the key assumptions in the considered system-level HetNet scenario. 
In Section \ref{sec:TDICIC}, the proposed time-domain dynamic RL based ICIC procedure is introduced. Additionally, a satisfaction equilibrium based time-domain ICIC technique enabling BSs to guarantee a minimum QoS level is presented. Section \ref{sec:FDICIC} presents the proposed dynamic RL based ICIC procedure in frequency-domain. In Section \ref{sec:SimulationResults}, the proposed solutions are validated in an  LTE-A system level simulator,  which is aligned with the simulation assumptions in 3GPP standardization studies~\cite{3GPP}, and Section \ref{sec:Conclusion} concludes the paper.


\section{System Model and Problem Formulation}\label{sec:SystemModelandProblemForml}

In this section, we present our system model and problem formulation for jointly optimizing the power allocation and traffic load among tiers. The goal of our learning based approaches in Section~\ref{sec:TDICIC} and Section~\ref{sec:FDICIC} is to develop strategies to solve the  optimization problem formulation presented in this section.

\subsection{System Model}\label{sec:SystemModel}
\begin{figure}
	\centering
		\includegraphics[width=0.5\textwidth]{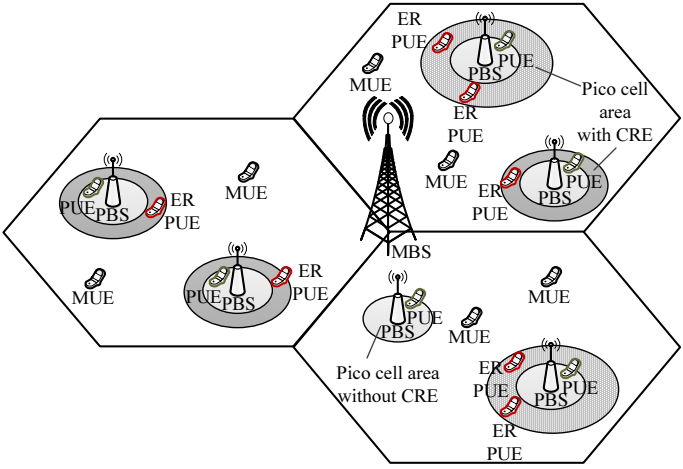}
	\caption{A heterogeneous scenario with cell range expansion (CRE).}
	\label{fig:HetnetScenario}
\end{figure}

We focus our analysis on a network deployment with multiple picocells overlaying a  macrocellular network consisting of three sectors per macrocell. A network consisting of a set of $\mathcal M=\{1,\ldots,M\}$ macrocells and a set of $\mathcal P=\{1,\ldots,P\}$ uniformly randomly distributed co-channel picocells per macro sector is considered, as depicted in Fig.~\ref{fig:HetnetScenario}. We consider that the total bandwidth (BW) is divided into subchannels with bandwidth $\Delta f = 180$ kHz. Orthogonal frequency division multiplexing (OFDM) symbols are grouped into resource blocks (RBs). Both macro- and picocells operate in the same frequency band and have the same number of available RBs, denoted by $R$. Without loss of generality, we consider that all transmitters and receivers have a single-antenna~\cite{MeryemMIMO}. A set of UEs $\mathcal U = \{1,\ldots,U\}$ is defined, whereby the UEs are dropped according to scenario \#4b in
 \cite{3GPP}, i.e. $\frac{2}{3}$ of UEs are uniformly dropped within a hotspot around picocells and the remaining UEs are uniformly dropped within the macrocellular area. All UEs (and BSs) are assumed to be active from the beginning of the simulations. We denote by $u(m)$ an MUE, while $u(p)$ refers to a PUE. We denote by 
$p^{m}_r(t_k)$ and $p^{p}_r(t_k)$ the downlink transmit power of MBS $m$ and PBS $p$ in RB $r$ at time instant $t_k$, respectively. Hereby, $t_k = kT_s$ is a time instant with $k = [1,\ldots,K]$, and $T_s$~=~1~ms. The SINR at an MUE $u$ allocated in RB $r$ of macrocell $m$ over one subframe duration, calculated over the subframe index $k$ is given by:
\begingroup\makeatletter\def\f@size{8}\check@mathfonts
\def\maketag@@@#1{\hbox{\m@th\large\normalfont#1}}
\begin{equation} \label{eq:SINR}
\gamma_r^{u}(t_k) = \frac{p_r^{m,u,\text{M}}(t_k)g_{m,u,r}^{\text{MM}}(t_k)}{\underbrace{\sum_{j = 1,j\neq m}^M p_r^{j,u,\text{M}}(t_k)g_{j,u,r}^{\text{MM}}(t_k)}_{I^\text{M}} + \underbrace{\sum_{p = 1}^{P} p_r^{p,u,\text{P}}(t_k)g_{p,u,r}^{\text{PM}}(t_k)}_{I^\text{P}} + \sigma^2}.\end{equation}
\endgroup
In \eqref{eq:SINR}, $g_{m,u,r}^{\text{MM}}(t_k)$ indicates the channel gain between the transmitting MBS $m$ and its MUE $u(m)$; $g_{j,u,r}^{\text{MM}}(t_k)$ indicates the link gain between the transmitting MBS $j$ and MUE $u$ in the macrocell at BS $m$;  $g_{p,u,r}^{\text{PM}}(t_k)$ indicates the link gain between the transmitting PBS $p$ and MUE $u$ of macrocell $m$; and $\sigma^2$ is the noise power. The interference terms caused by the MBSs and the PBSs are denoted by $I^\text{M}$ and $I^\text{P}$, respectively.

The SINR at an PUE $u$ allocated in RB $r$ of picocell $p$ over one subframe duration,  calculated over the subframe index $k$ is given by:
\begingroup\makeatletter\def\f@size{8}\check@mathfonts
\def\maketag@@@#1{\hbox{\m@th\large\normalfont#1}}
\begin{equation} \gamma_r^{u}(t_k) = \frac{p_r^{p,u,\text{P}}(t_k)g_{p,u,r}^{\text{PP}}(t_k)}{\underbrace{\sum_{j = 1,j\neq p}^P p_r^{j,u,\text{P}}(t_k)g_{j,u,r}^{\text{PP}}(t_k)}_{I^\text{P}} + \underbrace{\sum_{m = 1}^{M} p_r^{m,u,\text{M}}(t_k)g_{m,u,r}^{\text{MP}}(t_k)}_{I^\text{M}} + \sigma^2}.\end{equation}
\endgroup
In (2), $g_{p,u,r}^{\text{PP}}(t_k)$ indicates the link gain between the transmitting PBS $p$ and its PUE $u$; $g_{j,u,r}^{\text{PP}}(t_k)$ indicates the link gain between the transmitting PBS $j$ and PUE $u$ in the picocell at PBS $p$; and $g_{m,u,r}^{\text{MP}}(t_k)$ indicates the link gain between the transmitting MBS $m$ and PUE $u$ of PBS $p$.

In the scenario of Fig.~\ref{fig:HetnetScenario}, cell association is performed according to the maximum biased reference signal received power (RSRP) ~\cite{3GPP36839,Jeff1,Jeff3}. In particular, a UE-$u$  is handed over from cell $j$ to cell $l$ if the following condition is fulfilled:
\begin{equation}
P_{j,\text{RSRP}}(u) \text{ [dBm]} + \beta_j \text{ [dB]} < P_{l,\text{RSRP}}(u) \text{ [dBm]} + \beta_l \text{ [dB]},
\end{equation}
where $P_{j,\text{RSRP}}(u)$ ( $P_{j,\text{RSRP}}(u)$) is the $u$-th UE's RSRP from cell $j$ ( $l$) in dBm, and $\beta_j$ and $\beta_l$ are the range expansion bias of cell $j$ and $l$ in dB, respectively.
\subsection{Problem Fromulation}\label{sec:ProblemFormulation}
We focus on joint interference management and cell association in HetNets relying on both, the time- and frequency-domain ICIC mechanisms. Interference management is achieved by power control at both tiers, and cell association is optimized by REB $\beta^p$ adjustment per picocell. The considered optimization problem aims at achieving a target SINR for each UE $u(n)\in \mathcal{U}$ associated to BS $n$. The following joint power allocation and load balancing optimization problem formulation calculated over time instants $t_k$  for each BS $n$ is defined as follows:
\begin{align}\label{eq:Problemformulation}
\min_{\beta_{\rm lin}^p\atop p_r^n(t_k), p_r^m(t_k)} & \sum_{k=1}^K\sum_{u(n)\in \mathcal{U}}\vert \gamma_r^{u(n)}(t_k) - \gamma_{\rm target}\vert
\end{align}
\vspace{-0.2 cm}

\hspace{4.7cm}subject to:
\begin{align}
& \sum_{r=1}^R p_r^n(t_k) \leq p_{\text{max}}^n \hspace{0.5cm} \forall n\\
& \beta_{\text{lin}}^p = 10^{\beta^p/10} \text{, with } \beta^p \in \{0,  6, 12\} \text{ dB}
\end{align}
\vspace{-0.5 cm}

with $\sum_{r=1}^R p_r^n(t_k)= p_{\text{tot}}^n$ being the total transmit power of BS $n$, and the SINR after the biased cell association is $\gamma_r^{u(n)}(t_k) = \{\gamma_r^{u(m)}(t_k),\gamma_r^{u(p)}(t_k)\}$.

The optimization problem formulation in \eqref{eq:Problemformulation} aims at achieving a target SINR for each UE by joint power allocation and REB value adaptation for load balancing. Our system model focuses on a co-channel HetNet deployment, in which increasing the power level of a BS in one RB will cause interference to a UE scheduled on the same RB by another BS, so that the target SINR cannot be achieved by simply increasing the transmit power levels. Additionally, constraint (5) implies that the total transmit power of a BS is limited.


\section{Time-Domain ICIC: A Reinforcement Learning Perspective}\label{sec:TDICIC}

In this section, we first describe the time-domain ICIC approach in order to introduce our self-organizing learning procedures in time  domain. Our first approach leverages a dynamic reinforcement learning procedure in which picocells optimally learn their CRE bias in a heterogeneous deployment of picocells. Moreover, the macrocell learns which MUEs to schedule and on which RBs, while taking into account the picocell resource allocation. To do that, we consider a \emph{two-level} approach with loose coordination among macro and picocell tiers, in which the RAN autonomously optimizes the CRE bias value of picocells. At the same time the picocells dynamically learn their transmit power levels   to maximize the overall system performance. 

We propose a $Q$-learning formulation, which consists of a set $\mathcal P$ of PBSs and a set $\mathcal M$ of MBSs, denoted as the players/agents. We define a set of states $\mathcal{S}$ and actions $\mathcal{A}$ aiming at finding a policy that minimizes the observed costs over the interaction time of the players. Every player explores its environment, observes its current state $s$, and takes a subsequent action $a$ according to its decision policy $\pi:s\rightarrow a$. For all players, individual $Q$-tables maintain their knowledge of the environment to take autonomous decisions based on local and limited information. It has been shown that the $Q$-learning appraoch converges  to optimal values for Markov decision processes (MDPs) \cite{Harmon}, where the goal of a player is to find an \emph{optimal} policy $\pi^*(s)$ for each state $s$, so as to minimize the cumulative costs over time.

In some cases,  optimality is not aimed at,  and thus less complex algorithms are preferred, in which agents are solely interested in  guaranteeing a certain level of satisfaction to their users. Therefore, our second approach considers a satisfaction-based learning procedure based on game theory, which is a decentralized algorithm  allowing players to self-configure so as to achieve  satisfaction equilibria. This approach   guarantees that the QoS requirements  are satisfied in the network. The idea of satisfaction equilibrium was introduced in \cite{Ross,Ross2}, in which agents having partial or no knowledge about their environment are solely interested in the satisfaction of some individual performance constraints instead of individual performance optimization. Here, we consider a satisfaction  based game formulation that enables players (i.e., PBSs) to autonomously adapt their strategies to guarantee a certain level of QoS to UEs when optimality is not aimed for.

The main difference between $Q$-learning and satisfaction learning stems from the fact the former approach minimizes the total cost over time by trying different actions (trials and errors) as well as striking a balance between exploration and exploitation. As its name suggests, the latter algorithm guarantees that a given PBS does not update its strategy as long as its performance metric is satisfied. The rationale for using both algorithms is to underscore the tradeoffs of optimality vs. satisfaction. Rest of this section briefly summarizes the operation of classical time domain ICIC, and subsequently provides further details about the proposed $Q$-learning and satisfaction based learning time-domain ICIC and load balancing techniques. 

\subsection{Classical Time-Domain ICIC}

\begin{figure*}
	\centering
		\includegraphics[width=0.95\textwidth]{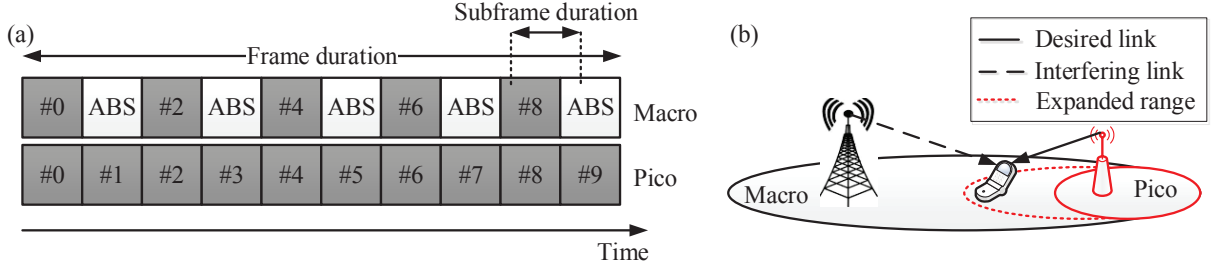}
		\caption{ Time-Domain ICIC in LTE-A with (a) Subframe structure, (b) Transmission by the picocell and the macrocell.}
	\label{fig:TDFDCRE}
\end{figure*}

The basic idea of time-domain ICIC is that an aggressor node (i.e. MBS) creates protected subframes for a victim node (i.e. PBS) by reducing its transmission power in certain subframes. These subframes are called Almost Blank Subframes (ABS). Notably, in co-channel deployments, ABSs are used to reduce interference created by  transmitting nodes while providing full legacy support. Fig. \ref{fig:TDFDCRE} depicts an ABS example with a duty cycle of $50\%$. During ABS subframes, BSs do not transmit data but may transmit reference signals, critical control channels, and broadcast information. For the example scenario in Fig. \ref{fig:TDFDCRE}, if the PBS schedules its PUEs which have low SINRs in subframes $\#1,\#3,\#5,\#7,\#9$, it protects such PUEs from strong inter-cell interference.

\subsection{$Q$-Learning based Time-Domain ICIC}
For the problem formulation of $Q$-learning, we divide the problem into a bias value selection and power allocation sub-problems. These two sub-problems are inter-related in which each picocell, as a player, individually selects first a bias value for CRE by considering its own PUEs' QoS requirements, after which the transmit power is optimally allocated. Additionally, we consider the MBS as a second type of player, which performs learning after the picocell has selected its bias values and transmit power levels per RB. We name this learning approach as \emph{dynamic} $Q$-learning, in which the picocell  informs the MBS  which RBs are used for scheduling  ER PUEs through  the X$2$ interface. These RBs will be protected by the MBS by using lower power levels. In case of more than one PBS, the MBS considers the protected RBS of all PBSs, and optimizes its transmit power allocation on these protected RBs as well as on the remaining RBs. Formally speaking, the player, state, action and perceived cost associated to the  Q-learning procedure are defined as follows:

\begin{itemize}
\item \textbf{Player:} PBS $p, \forall 1\le p\le P$ and MBS $m, \forall 1\le m\le M$.
\item \textbf{State:} The state representation of player $n$ at time $t_k$ in RB $r$ is given by the vector state $\vec s_r^{\hspace{0.1cm} n} = \{I_r^{u(p)},I_r^{u(m)}\}$.

\begin{equation} I_r^{u(n)}=\begin{cases} 0, & \text{if} \hspace{0.2cm}\Gamma_r^{u(n)} < \Gamma_{\text{target}} - 2 \text{ dB}\\
                                                    1, & \text{if} \hspace{0.2cm} \Gamma_{\text{target}} - 2 \text{ dB} \leq \Gamma_r^{u(n)} \leq \Gamma_{\text{target}} + 2 \text{ dB}\\
                                                    2, & \text{otherwise}   \end{cases},\end{equation}
where $n = \{p,m\}$, $\Gamma_r^{u(n)}=10\log\left(\gamma_r^{u(n)}\right)$ is the instantaneous SINR of UE $u$ in dB in RB $r$, and $\Gamma_{\text{target}}=20$ dB  is the target SINR value. In our state definition, we consider both MUE and PUE interference levels, which implies that both players optimize both type of UEs' states. We consider a target SINR of 20~dB and define a range within which the instantaneous SINR is satisfied. This range is selected to be small, i.e., $\pm2$~dB~\cite{Mehlfuhrer}, to be close to the target SINR. The main motivation for defining such a range is that it is very difficult to maintain exact SINR values for each of the UEs at each BS.  In particular, even when a UE's SINR is very close to the target SINR, if an exact SINR  is aimed, the UE will be considered not to be in the targeted state, hence yielding stability problems. We consider the range of $\pm2$~dB to be ‘acceptable’, because we target a BLER of $10\%$, which is required for LTE systems~\cite{36213}. According to our link-to-system level mapping look-up table, this target BLER still holds for the second largest CQI value 14. In this case, the SINR decreases by 2 dB. Hence, we selecta  2 dB degradation as an ‘acceptable’ range for the target SINR and since the cost function is parabolic, we consider a symmetric range of $\pm2$~dB.

\item \textbf{Action:} For player $\text{PBS}$, the action set is defined as $A^p = \{\beta^p,a_r^p\}_{r \in \{1,...,R\}}$, where $a_r^p$ is the transmit power level of PBS $p$ over a set of RBs $\{1,...,R\}$, and $\beta^p$ is the bias value for CRE of PBS $p$. It has to be pointed out that the bias value setting will influence the convergence behavior of the learning algorithm and that the presented bias values have been selected experimentally.
For player $\text{MBS}$, the action set is defined as $A^m = \{a_r^m\}_{r \in \{1,...,R\}}$, where $a_r^m$ is the transmit power level of MBS $m$ over a set of RBs $\{1,...,R\}$. Different power levels are defined for protected RBs. 
\item \textbf{Cost:} The considered cost in RB $r$ of player $n$ is given by
\begin{equation}\label{cost}
 c_r^n = \begin{cases} 500, & \text{if} \hspace{0.2cm} P_{\text{tot}}^n > P_{\text{max}}\\
 (\Gamma_r^{u(n)} -\Gamma_{\text{target}})^2 , & \text{if} \hspace{0.2cm} \text{otherwise}\\
 \end{cases}.
\end{equation}
 The rationale behind this cost function is that the $Q$-learning aims to minimize its cost, so that the SINR at UE $u$ is close to a selected target value $\Gamma_{\text{target}}$. Considering a cost function with a minimum as target SINR  as in \eqref{cost}, will enable the player to develop a strategy that leads to SINR values close to the target SINR. The target SINR is set to be 20 dB, and this corresponds to a maximum CQI level of 15 in typical look-up tables~\cite{Mehlfuhrer}. Therefore, setting an SINR target of $20$~dB is considered as a reasonable optimization goal for the proposed $Q$-learning approach. The considered cost of 500 is only for the case that the total transmit power $P_{\text{tot}}^n$ is larger than the maximum transmit power of a BS. It provides the best performance and convergence trade-off in our simulations as shown in Section~\ref{sec:Convergence} and has been heuristically selected~\cite{Ana}.
\end{itemize}

Being in state $s$ after selecting action $a$ and receiving the immediate cost $c$, the agent updates its knowledge $Q(s,a)$ for this particular state-action pair as follows:
\begin{equation}\label{eq:Qupdate}
Q^n(s,a)\leftarrow (1-\alpha)Q^n(s,a)+ \alpha[c^n+\lambda \min_a Q^n(s',a)],\end{equation}
where $\alpha = 0.5$ is the player's willingness to learn from its environment, $\lambda = 0.9$ is the discount factor, and $s'$ is the next state~\cite{Ana,Meryem}. 
Hereby, the agent's previous knowledge about the state-action pair $(s,a)$ is represented by the first term in \eqref{eq:Qupdate}. On the other hand, the second term represents the agent's learned value, which consists of the received cost $c^n$ after executing action $a$ and the estimated minimum future cost $\min_a Q^n(s',a)$. Hence, $Q$-learning is an iterative procedure in which the previous knowledge ($Q^n(s,a)$) is updated by considering the newly obtained knowledge represented by the cost value $c$ and estimates of future costs $\min_a Q^n(s',a)$.

In addition to the  $Q$-learning formulation, referred to as \emph{dynamic QL} in the following, we also consider the scenario where there is only one player: the PBS. In this approach, only the PBS is carrying out the decentralized learning procedure, and informs the MBS about the RBs allocated to ER PUEs to be considered as ABSs. Subsequently, the MBS  uses those ABS patterns on these RBs and  uniformly distributes its transmit power over the remaining RBs. Through the rest of the paper, this variation of the $Q$-learning formulation is refered as \emph{static QL}.

\subsection{Satisfaction Based Learning in Time-Domain ICIC}\label{sec:satisfaction}

As discussed before, the $Q$-learning based ICIC procedure aims at optimality by achieving a target SINR for the MUEs, we propose another approach that guarantees a level of QoS satisfaction. This approach does not achieve the target SINR values as defined for the $Q$-learning based ICIC procedure, however, it is less complex than the $Q$-learning based approach in terms of memory  and computational requirements. Compared to $Q$-learning the agents do not have to store a table reflecting their knowledge for each state-action combination. Instead, a probability distribution over all actions is stored. Hence, instead of $|\mathcal{S}|\times |\mathcal{A}|$ only $1\times |\mathcal{A}|$ information is stored in the satisfaction based learning. A discussion about the memory and computational requirements of both approaches is presented in Appendix B.

The satisfaction based learning algorithm is defined as a game in satisfaction-form
\begin{equation} \mathcal G = \{\mathcal P, \{\mathcal A_p\}_{p \in \mathcal P}, \{u_p\}_{p \in \mathcal P}\}.\end{equation}

The set $\mathcal A_p = \{A_p^{(1)},\ldots,A_p^{(N_p)}\}$ represents the set of $N_p$ actions PBS $p$ can select. An action profile is a vector $a = (a_1,\ldots,a_P)\in A$, where $\mathcal A = \mathcal A_1 \times \ldots \times \mathcal A_P$. For all $p \in \mathcal P$, the function $u_p: \mathcal A \rightarrow \mathbb R_+$ is the utility function of PBS $p$ (see definition in \eqref{utility} for time-domain ICIC algorithm at time $t_k$).

We decompose our satisfaction based learning algorithm into two inter-related sub-problems. The PBS first selects a bias value for CRE by considering its own PUEs' QoS requirements. Subsequently, it selects the transmit power on RB $r$ according to a discrete probability distribution $\pi^p_{r,n_p}(t_k)= (\pi^p_{r,1}(t_k), \ldots, \pi^p_{r,|\mathcal A_{p}|}(t_k))$. Here, $\pi^p_{r,n_p}(t_k)$ is the probability with which the PBS $p$ chooses  action $a_{r,n_p}^{p}(t_k)$ on RB $r$ at time instant $t_k$, which are the same power levels as in the $Q$-learning algorithm. And, $n_p \in N_p \triangleq \{1, \ldots, |\mathcal A_p|\}$ is the element's index of each set $\mathcal A_p, \forall p\in P$.
We define player $p$'s utility function $u_p(t_k)$ at time instant $t_k$ as the achievable rate
\begin{equation}\label{utility}
u_p(t_k) = \sum_{r\in R} \log_2(1+\gamma_r^{u(p)}(t_k)).
\end{equation}

The proposed satisfaction-based time-domain ICIC technique is carried out as follows. First, at time instant $t_k = 0$,  each player $p$ sets its initial probability distribution $\pi^p_{n_p}(0)$\footnote{For brevity, let the RB index $r$ be dropped from the formulation in the sequel}, and selects its initial action $a_{n_p}^{p}(0)$ following an arbitrary chosen probability distribution per RB $r$. Subsequently, at time instant $t_k > 0$, each player chooses its action $a_{n_p}^p(t_k)$ according to its probability distribution $\pi^p_{n_p}(t_k)$. This probability distribution is updated if the target utility $u_{\text{target}}=\sum_{r\in R}\log_2\left(1+10^{\frac{\Gamma_{\text{target}}}{10}}\right)$ is not achieved, following the step size of probability updating rule. For the considered problem formulation, the step size is given by:

\begin{equation}\label{eq:b(t)}
 b_p(t_k) = \dfrac{u_{\text{max},p}+ u_{p}(t_k)-u_{\text{target}}}{2u_{\text{max},p}},
 \end{equation}
where ${u}_{p}(t_k)$ is the observed utility and $u_{\text{max},p}$ is the highest utility the PBS $p$ can achieve in a single player scenario. Subsequently, every PBS $p$ updates its action $a^p_{n_p}(t_k)$ at each time $t_k$ according to a probability update function $d_p(\pi^p_{n_p}(t_k))$, which is defined as follows:
\begingroup\makeatletter\def\f@size{8}\check@mathfonts
\def\maketag@@@#1{\hbox{\m@th\large\normalfont#1}}
\begin{equation}\label{eq:ProbUdpate}
d_p(\pi^p_{n_p}(t_k)) = \pi^p_{n_p}(t_k) + \tau_p(t_k) b_p(t_k) \left(  \mathds{1}_{\{a^p_{n_p}(t_k)= a^p_{n_p}\}} - \pi_{n_p}(t_k)\right),
\end{equation}\endgroup
where $\forall p\in P,\tau_p(t_k) = \frac{1}{t_k+T_s}$ is the learning rate of the PBS $p$. The rationale behind this probability update function is to update the probability of selecting action $a_{n_p}^p(t_k)$ based on the step size $b_p(t_k)$ in \eqref{eq:b(t)}, which is a function of the observed utility.

If the observed utility $u_p(t_k)$ is larger than the target utility, i.e. if the agent is satisfied, the PBS selects the same action as at time $t_k-T_s$ as described in the first condition of \eqref{eq:ActionSelection}. Otherwise it selects the action according to the probability distribution function $\pi^p_{n_p}(t_k)$, as follows:

 \begin{equation}\label{eq:ActionSelection}
a^p_{n_p}(t_k)= \begin{cases} a^p_{n_p}(t_k-T_s), & \text{if } {u}_{p}(t_k)\geq u_{\text{target}}\\
a^p_{n_p}(t_k) \sim \pi^p_{n_p}(t_k)& \text{otherwise}\\
\end{cases},
\end{equation}
where $\sim$ means according to the probability distribution $\pi_{n_p}^p(t_k)$. The probability distribution is then updated as follows:

\begin{equation}\label{eq:ProbUpdate}
\pi^p_{n_p}(t_k) = \begin{cases} \pi^p_{n_p}(t_k), & \text{if } u_{p}(t_k)\geq \Gamma_{\text{target}} \\
d_p(\pi^p_{n_p}(t_k-T_s))& \text{otherwise}\\
\end{cases}.
\end{equation}

Finally, this learning procedure is repeated until convergence, which is proven based on the following proposition. 

\emph{Proposition - 1:} The behavioral rule in equation \eqref{eq:ActionSelection}-\eqref{eq:ProbUpdate}  with probability distributions \\ ${\pi^p_{r,n_p}(t_k)= (\pi^p_{r,1}(t_k), \ldots, \pi^p_{r,|\mathcal A_{p}|}(t_k))\in \mathcal A_p}$, with $p \in \mathcal P$, converges to an equilibrium of the game  ${\mathcal G = \{\mathcal P, \{\mathcal A_p\}_{p \in \mathcal P}, \{u_p\}_{p \in \mathcal P}\}}$ in finite time if for all $p \in \mathcal P$ and for all $n_p \in \{1, \ldots, |\mathcal{A}_p|\}$, it holds that
$\pi_{r,n_p}^p(t_k) > 0$.

\emph{Proof:} See Appendix A. $\blacksquare$

\section{Frequency-Domain ICIC: A Reinforcement Learning Perspective} \label{sec:FDICIC}

In this section, after describing the classical frequency-domain ICIC as defined in 3GPP, we introduce new frequency domain ICIC and load balancing algorithms based on RL techniques. In contrast to existing frequency-domain ICIC solutions like single-flow CA (where PBSs select one CC and apply a fixed CRE bias), we consider a heterogeneous case where  different CRE bias values are used across different CCs in a self-organizing manner. In such a scenario, we formulate  dynamic frequency-domain ICIC approaches applied both, to single and multi-flow CA settings. On the other hand, the $Q$-learning based ICIC is considered in a similar way as it was discussed for the time-domain ICIC.

\begin{figure*}
	\centering
		\includegraphics[width=0.95\textwidth]{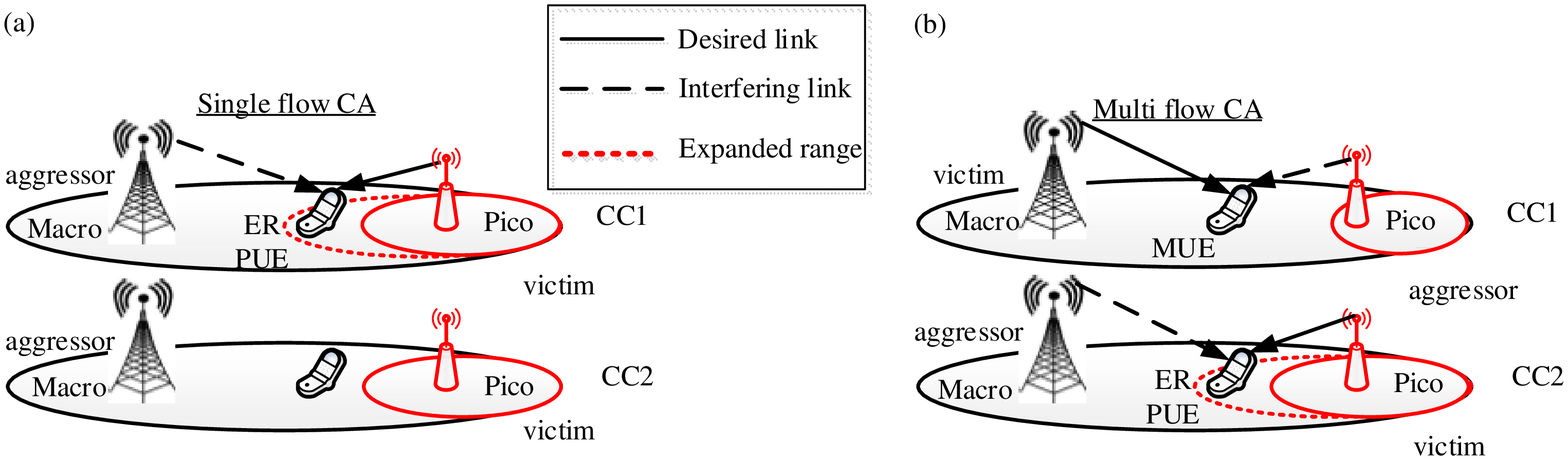}
		\caption{Scheduling of an ER PUE (in case of CRE)/cell-edge MUE (in case of no CRE) in frequency-domain ICIC with (a) single-flow and (b) multi-flow CA.}
	\label{fig:MFSF}
\end{figure*}

\subsection{Classical Frequency-Domain ICIC}

In 3GPP Release 12, frequency-domain ICIC is performed through the concept of CA.  In \cite{Jeff}, CA is studied as a function of bias values and frequency band deployment, in which CA enables UEs to connect to several carriers simultaneously. Two different methods are considered, namely the single- and multi-flow CA. In single-flow CA, the MBS is the aggressor cell and the PBS is the victim cell as depicted in Fig. \ref{fig:MFSF} (a). The PBS performs CRE on CC1 to offload the macrocell and serves its ER PUE on this CC, so that the MBS is the interfering BS in CC1. In CC2, the PBS does not perform CRE, so that the ER PUE is only served on CC1 and the remaining PUEs can be served on CC2. Hence, single-flow CA enables UEs to connect to one BS at a time. 

A recent feature in 3GPP Release-12, referred to as multi-flow CA, enables a better use of resources and improves system capacity. As depicted in Fig. \ref{fig:MFSF} (b), in multi-flow CA multiple BSs (from different tiers)  simultaneously transmit data to a UE on different CCs \cite{3GPP25211,3GPP25212,3GPP25213,3GPP25214}.  While the MBS remains still the aggressor cell on  CC1, in which PBS perform CRE, it becomes the serving cell on CC2. Hence, in single-flow CA, UEs associate with only one of the available tiers at a given time and in multi-flow CA based HetNets,  UEs can be served by both macro- and picocells at the same time. This necceciates a smart mechanism in which the different tiers coordinate their transmission through adaptive cell range expansion across different CCs.

\subsection{Dynamic Frequency-Domain ICIC for Single-Flow CA}

We divide the single-flow CA problem into primary CC selection, bias value selection and power allocation sub-problems. These three sub-problems are inter-related in which the PBS and MBS (as players) learn their optimal ICIC strategy, which is presented in Algorithm I. The PBS first selects its optimal CC to perform CRE, then the bias value for CRE in the selected CC, after which the transmit power is allocated accordingly. Hence, we consider a three-stage decision making process, in which the MBS is informed about the PBS's primary CC via the X$2$ interface. The MBS selects PBS's secondary CC as its primary CC and learns its optimal power allocation strategy. In a network with more than one PBS and more than one CC, each PBS may select different CCs as their primary CC. In this case, we propose that the MBS selects that CCs as its primary CC, which has been selected by less number of PBSs. In case of equality, the CC which will lead to larger performance degradation caused my MBS interference is selected. While MBS selects low power levels on its secondary (PBS's primary) CC, it selects higher power levels on its primary CC. The rationale behind considering two different power levels for MBS's primary and secondary CC, is to reduce interference on ER PUEs, which are served on PBS's primary CC. The main difference with the dynamic time domain ICIC learning procedure discussed in Section IV is in the action definition. Hence, we redefine our action formulation as follows:
\begin{itemize}
\item \textbf{Action:} For player PBS $p$  the action set is defined as, $A^p = \{C_i^p,\beta^p,a_r^p\}_{r \in \{1,...,R\}}$,
where $C^p_{i_{i\in \{1,2\}}}$ is the selected component carrier to perform CRE on the selected CC, $\beta^p\in\{0,6,12\}$ dB is the bias value for CRE on selected $C_i^p$ of PBS $p$ and $a_r^p$ is the transmit power level of PBS $p$ over a set of RBs $\{1,...,R\}$. Hence, the PBSs will independently learn which CC it performs range expansion, with which bias value, and how to optimally perform power allocation.\\
For player MBS $m$ the action set is defined as, $A^m = \{a_{r,C_i}^m\}_{r \in \{1,...,R\}}$, where $a_{r,C_i}^m$ is the transmit power level of MBS $m$ over a set of RBs $\{1,...,R\}$ on CC $C_i$. Different power levels are defined for MBS's primary and secondary CCs.
\end{itemize}
\begin{algorithm}[t]
\small\caption{Dynamic $Q$-learning based ICIC algorithm for single-/multi-flow CA.} \label{alg:SingleflowCA}
\begin{algorithmic}[1]\vspace{1ex}
\LOOP
\FOR{player $p$}
\STATE Select primary CC $C^p\in\{1,2\}$
\STATE Select bias value $b^p$ for primary CC $C^p$
\STATE Select power level $a_r^p$ according to $\arg\min_{a\in \mathcal{A}^p}Q^p(s,a)$ on both CCs
\ENDFOR
\STATE Inform player $m$ about primary CC $C^p$
\FOR{player $m$}
\STATE Select player $p$'s secondary CC as primary CC $C^m$
\IF{multi-flow CA}
\STATE Select bias value $b^m$ for primary CC $C^m$
\ENDIF
\STATE Select power level $a_r^m\in \mathcal{A}^m$ according to $\arg\min_{a\in \mathcal{A}^m}Q^m(s,a)$
\ENDFOR
	  \STATE Receive an immediate cost $c$
    \STATE Observe the next state $s'$
    \STATE Update the table entry according to equation \eqref{eq:Qupdate}
    \STATE $s=s'$
\ENDLOOP
\end{algorithmic}
\end{algorithm}
\subsection{Dynamic Frequency-Domain ICIC for Multi-Flow CA}
In contrast to the single-flow CA in which the MBS is always the aggressor cell, in multi-flow CA either the MBS or the PBS is the aggressor cell. This is because both MBS and PBS perform CRE on their primary CCs, so that a UE can be served on different CCs by different BSs based on its biased received power. Similar to the single-flow CA learning algorithm, the multi-flow CA based ICIC learning algorithm assumes PBS and MBS as players. The main difference with the single-flow CA based ICIC learning algorithm, is the action definition, which is highlighted in the IF-condition in line 11 of Algorithm \ref{alg:SingleflowCA}.
\begin{itemize}
\item \textbf{Action:} For player PBS $p$  the action set is defined as, $A^p = \{C_i^p,\beta^p,a_r^p\}_{r \in \{1,...,R\}}$, and for player MBS $m$ the action set is defined as, $A^m= \{C_i^m,\beta^m,a_r^m\}_{r \in \{1,...,R\}}$,
where $C_{i_{i\in \{1,2\}}}$ is the component carrier index that can be selected in order to perform CRE on the selected CC, $\beta\in\{0,6,12\}$ dB is the bias value for CRE on selected CC $C_i$, and $a_r$ is the transmit power level over a set of RBs $\{1,...,R\}$. Hence, the PBSs and MBS will independently learn which CC they perform range expansion, with which bias value and how to optimally perform the power allocation. Since, both PBS and MBS can be aggressor cells, different power levels are considered for CCs on which the BSs perform CRE, and the regular CCs which do not have CRE. 

In addition, we consider the case of one player formulation, in which the PBS is the player. In this case, PBS carries out the multi-flow CA based $Q$-learning procedure and informs MBS about its primary CC and MBS uses reduced power levels on this CC. However, even if no CRE is performed by the MBS, a UE can be served by both PBS and MBS on different CCs at the same time. This learning algorithm will be coined as \emph{MF static QL} while the two player algorithm  is named \emph{MF dynamic QL}.
\end{itemize}

\begin{table*}
\begin{center}
\caption{Simulation parameters.}
\begin{tabular}{|l|l||l|l|}\hline
\textbf{Parameter} & \textbf{Value} & \textbf{Parameter} & \textbf{Value}\\\hline
Cellular layout & Hexagonal grid, & Transmission mode & Transmit diversity\\
& 3 sectors per cell & &\\\hline
Carrier frequency & 2 GHz & Macro path loss model & $128.1 + 37.6\log_{10}(R)$ dB ($R$[km])\\\hline
System bandwidth & 10 MHz & Pico path loss model & $140.7 + 36.7\log_{10}(R)$ dB ($R$[km])\\\hline
Bandwidth per CC & 5 MHz & Traffic model & Full buffer\\\hline
Subframe duration & 1 ms & Scheduling algorithm & Proportional fair \\\hline
Number of RBs & 50 & MUE speed & $ 3\frac{\text{km}}{\text{h}}$\\\hline
Number of macrocells & 1 & Min. dist. MBS-PBS & 75 m\\\hline
Number of PBSs per macrocell $P$ & \{2,4,8\} & Min. dist. PBS-PBS & 40 m\\\hline
Max. macro (pico) BS  & $P_{\text{max}}^M = 46$ dBm & Min. dist. MBS-MUE & 35 m\\
transmit power & ($P_{\text{max}}^P = 30$ dBm) & & \\\hline
Number of UEs per sector $N_{\text{UE}}$& 30 & Min. dist. PBS-PUE & 10 m\\\hline
Number of hotspot UEs $ N_{\text{hotspot}}$ & $\lceil 2/3 \cdot N_{\text{UE}}/P \rceil$ & PUE radius & 40 m\\\hline 
Thermal noise density & -174 dBm & Macro (Pico) antenna gain & 14 dBi (5 dBi)\\\hline
\end{tabular}
\end{center}
\end{table*}

\begin{figure*}[htb!]
\centering
\includegraphics[width=0.95\textwidth]{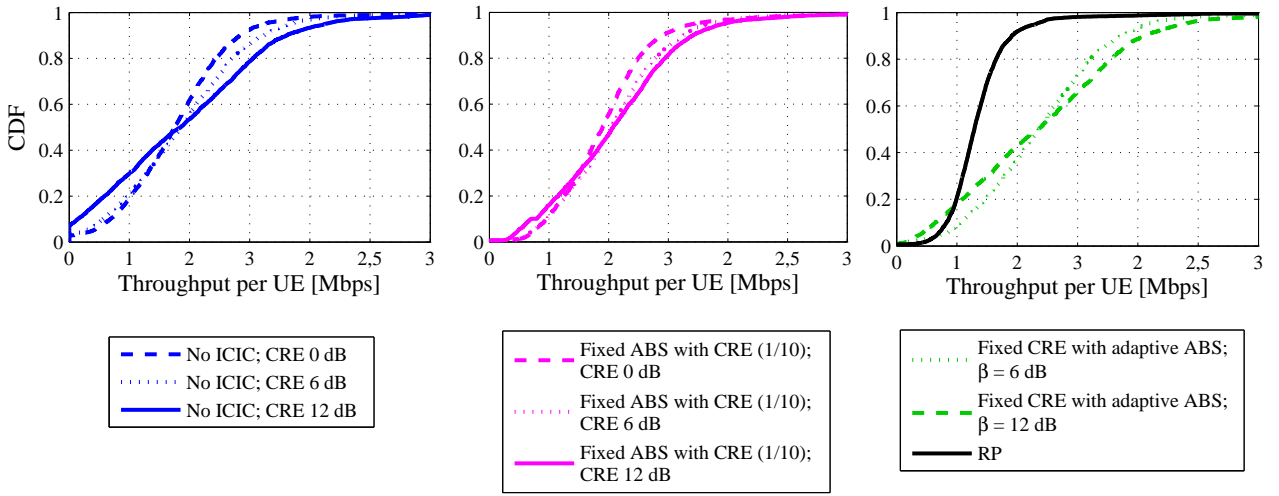}
\caption{CDF of the UE throughput of the reference algorithms in time-domain ICIC.}
\label{fig:TimeDomainRefs}
\end{figure*}


\section{Simulation Results}\label{sec:SimulationResults}
In this section, the proposed solutions are validated in a 3GPP-compliant LTE-A system-level simulator. First, time-domain ICIC results are discussed followed by frequency domain ICIC results.
The system-level simulator is based on snapshots, i.e. in each iteration the transmission time interval (TTI) of 1 ms is simulated~\cite{MeryemSimulator}. All system layout, channel model and  BS assignment methods are based on 3GPP configurations\cite{3GPP}. A time- and frequency selective channel is considered with shadowing correlation of 0.5 between cells and shadowing standard deviation of 8 dB. The user association is based on the strongest (biased) reference signal received power (RSRP).

The scenario used in our system-level simulations is based on configuration $\#4$b in \cite{3GPP}. We consider a macrocell consisting of three sectors and $P = \{2,4,8\}$ PBSs per macro sector, uniformly randomly distributed within the macrocellular environment. $N_{\text{UE}} = 30$ mobile users are generated within each macro sector from which $N_{\text{hotspot}} = \lceil \frac{2}{3} \cdot N_{\text{UE}}/P\rceil$ are randomly and uniformly dropped within a 40 m radius of each PBS. The remaining UEs are uniformly distributed within the macrocellular area. All UEs have an average speed of $3$ km/h. A full buffer traffic model is assumed. Without lost of generality, we do not consider any (feedback) delays throughout the simulations due to computational limitations. Since a velocity of $3$ km/h is assumed, the channel conditions do not change significantly within milliseconds, so that the shape of the presented results will remain the same/similar if delays are considered. Further details about the system level simulation parameters are provided in Table I.

\subsection{Benchmark Solutions}
For the performance comparison of our proposed self-organizing  solutions, the following benchmark references are considered:
\begin{itemize}
\item \emph{Resource Partitioning (RP):} The MBS and the PBSs uniformly distribute their transmit powers among RBs. Half of the RBs are used by the macrocell, and the other half is reused by the picocells. This way, cross-tier interference is avoided \cite{Guvenc}.
\item \emph{No ICIC with CRE:} Cell range expansion is performed without any inter-cell interference coordination. Here, a bias of $\beta = [0;6;12]$ dB is added to the UE's the DL received signal strength by PBSs;  $\beta = 0$ dB means  no CRE.
\item \emph{Fixed ABS with CRE:} An ABS ratio of $\{1/10,3/10,7/10\}$ with CRE is considered, which describes the ratio between ABS and the total number of downlink subframes in a frame, i.e. TTIs in which the MBS does not transmit. The PBS transmits with uniform power allocation over all RBs in all TTIs.
\item \emph{Fixed CRE with adaptive ABS}: Uniform power allocation is performed using fixed CRE bias values for each picocell. Using an X2 interface, the MBS is informed on which RBs the ER PUEs are scheduled. The MBS mutes only on these RBs, so that they define the ABS pattern for the MBS.
\end{itemize}

\begin{figure}
\centering
\includegraphics[width=0.5\textwidth]{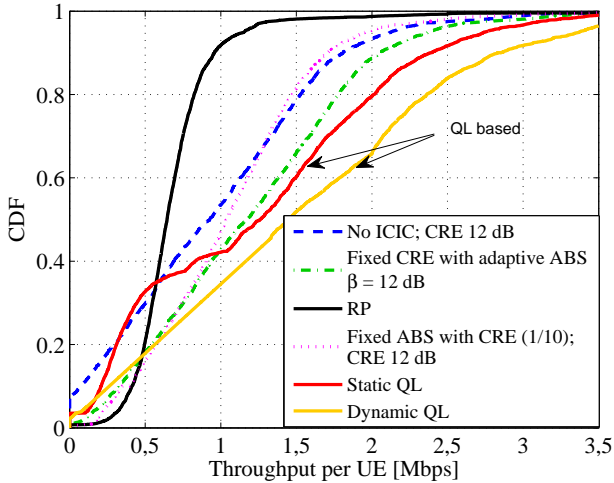}
\caption{CDF of the UE throughput of the $Q$-learning based algorithms in comparison with reference algorithms in  time-domain ICIC.}
\label{fig:TimeDomainResults}
\end{figure}

System level simulation results in terms of average UE throughput are presented in Fig. \ref{fig:TimeDomainRefs} and Fig. \ref{fig:TimeDomainResults} for time domain ICIC. The throughput values are obtained based on exponential effective SINR mapping (EESM) and look-up tables for the link level abstraction. No further link level protocols are considered for the evaluation of the presented system level simulation results.
The Cumulative distribution functions (CDFs) are plotted for a scenario, in which $2$ picocells per macrocell are activated. To provide a better overview, we split the CDF of the reference algorithms (except RP) and the proposed $Q$-learning based algorithms together with the  average best reference algorithms (and RP) into two figures. 

In Fig. \ref{fig:TimeDomainRefs}, it can be observed that increasing the CRE bias values without any inter-cell interference coordination results in very low data rates for cell-edge PUEs. The reason is two-fold. On the one hand, UEs select the picocells even though they are not the strongest cell and hence suffer from interference from MBS since the received signal of MBS is larger than that of the picocells.

On the other hand, PBSs may allow too many handovers and may not be able to guarantee QoS requirements of their own PUEs. A \emph{fixed CRE with adaptive ABS} with $\beta = 6$ dB is good for cell edge UEs whereas \emph{fixed CRE with adaptive ABS} with $\beta = 12$ dB is detrimental as the ER PUEs get exposed to MBS interference; the latter  is better for UEs with good channel conditions (i.e., higher percentiles). The \emph{fixed ABS with CRE $1/10$} clearly results in bad performance for each bias value for  UEs due to resource under-utilization. Yet, a larger performance degradation is seen in the \emph{RP} scheme (due to resource under-utilization).

The \emph{static QL} learning scheme in which PBSs learn how to select their optimal beta values and transmit power levels, and where MBSs use static ABSs, achieves high data rates while  yielding poor performance for cell-edge UEs, as shown in Fig. \ref{fig:TimeDomainResults}. Our proposed \emph{dynamic QL}  approach, in which PBSs perform $Q$-learning by considering their own PUEs' QoS requirements yields the best performance; PBSs do not increase the beta values,  without considering QoS requirements of their PUEs. On average, we obtain a gain of $125\%$ compared to the \emph{RP} and  $23\%$ compared to the \emph{Fixed CRE} with $\beta = 12$ dB.  The rationale is that the PBS informs the MBS which RBs are used for scheduling its ER PUEs, leading  the MBS  to reduce its power levels. Ultimately, the highest data rates are achieved by the proposed \emph{dynamic} QL approach while not being worse than any of the reference scheme. Finally, it is worth noting that the proposed self-organization approach hinges on a loose coordination in the form of RB indices used for ER PUEs' among macro and picocell tiers, and depending on traffic load  every PBS adopts a different CRE bias value so as to optimize its serving QoS requirements.
\subsection{ABS Power Reduction}
\begin{figure}[t]
\centering
   \begin{subfigure}[b]{0.45\textwidth}
   \includegraphics[width=1\textwidth]{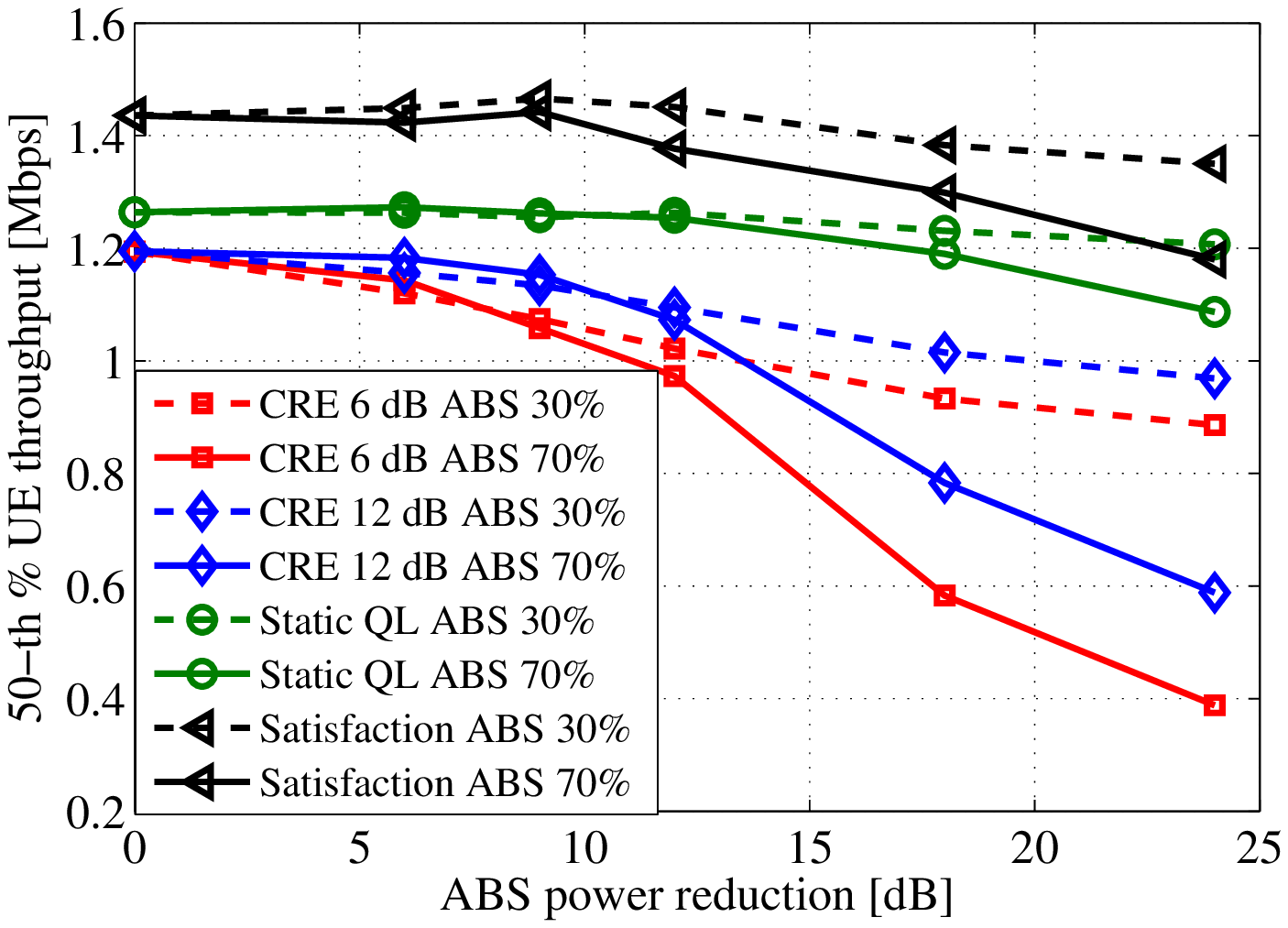}
  \put(-230,170){(a)}
\end{subfigure}
\begin{subfigure}[b]{0.45\textwidth}
   \includegraphics[width=1\textwidth]{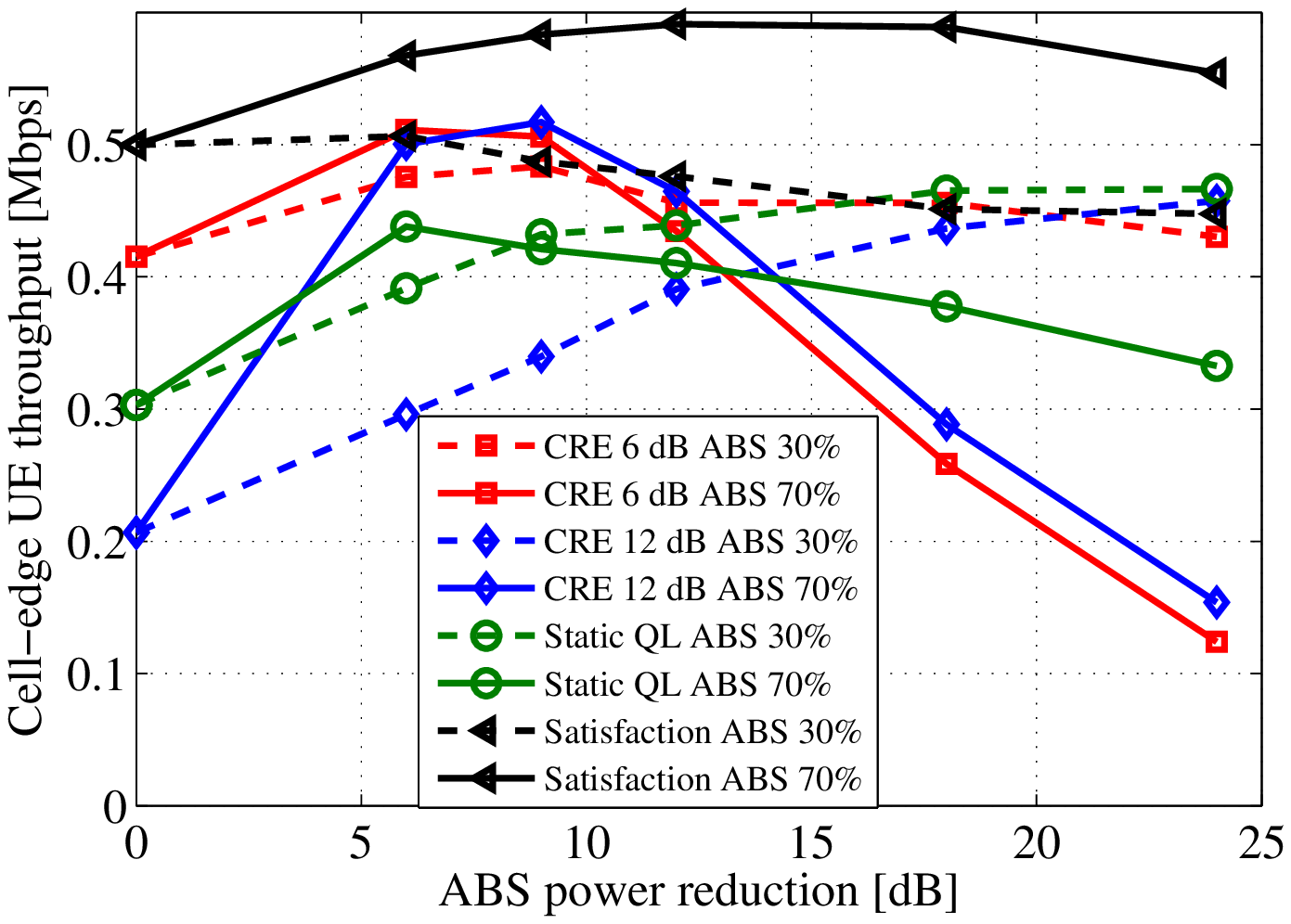}
  \put(-230,170){(b)}
\end{subfigure}
\caption{(a) 50-th \% UE throughput as a function of the ABS power reduction in time domain ICIC, and (b) Cell-edge UE throughput  as a function of the ABS power reduction in time domain ICIC.}
\label{fig:CelledgeResults}
\end{figure}

We also evaluate the performance of our proposed time domain algorithms for the ABS ratios 3/10 and 7/10 with reduced MBS transmission power in a HetNet scenario consisting of 2 picocells per macro sector. The ABS ratio describes  the ratio between subframes in which the MBS mutes and regular downlink subframes in which transmission is performed by MBS. 
We plot in Fig. \ref{fig:CelledgeResults} (a) and (b) the, 50-th \% and 5-th \% UE throughput performance versus the ABS power reduction for the static algorithms with \emph{fixed CRE} of 6 dB and 12 dB,  $Q$-learning based and satisfaction based ICIC schemes. In all simulations, the MBS transmission power reduction in ABS is $\{0, 6, 9, 12, 18, 24 \}$ dB.

Fig. \ref{fig:CelledgeResults} (a) plots the 50-th\% UE throughput. 
It can be seen that a mix of CRE bias values among picocells yields always  better average UE throughput in both ICIC techniques.
It is observed that low power ABS reduces the sensitivity to ABS ratio. Reduced ABS ratio sensitivity is also observed in the satisfaction based ICIC algorithm for low power ABS, whereas the $Q$-learning based ICIC technique is almost insensitive to the ABS ratio for all ABS power reduction values.

The 5-th\% UE throughput results are shown in Fig. \ref{fig:CelledgeResults} (b). For the \emph{fixed CRE} technique it can be observed that there exists an optimum ABS power setting for each combination of CRE bias and ABS ratio; which is 6 dB to 9 dB power reduction. The corresponding optimum ABS ratios are 3/10 for 6 dB CRE bias and 7/10 for 12 dB CRE bias. 
For a ABS ratio of 3/10, the proposed ICIC techniques perform very similarly, while for ABS ratio of 7/10 the satisfaction based ICIC algorithm outperforms all cases. 
Especially, in the optimum region of the \emph{fixed CRE} technique, the proposed learning algorithms cannot show any enhancement, except the satisfaction based ICIC, with ABS ratio 7/10. 

\begin{figure*}
\centering
\begin{minipage}{.29\textwidth}
  \centering
  \includegraphics[width=1.0\linewidth]{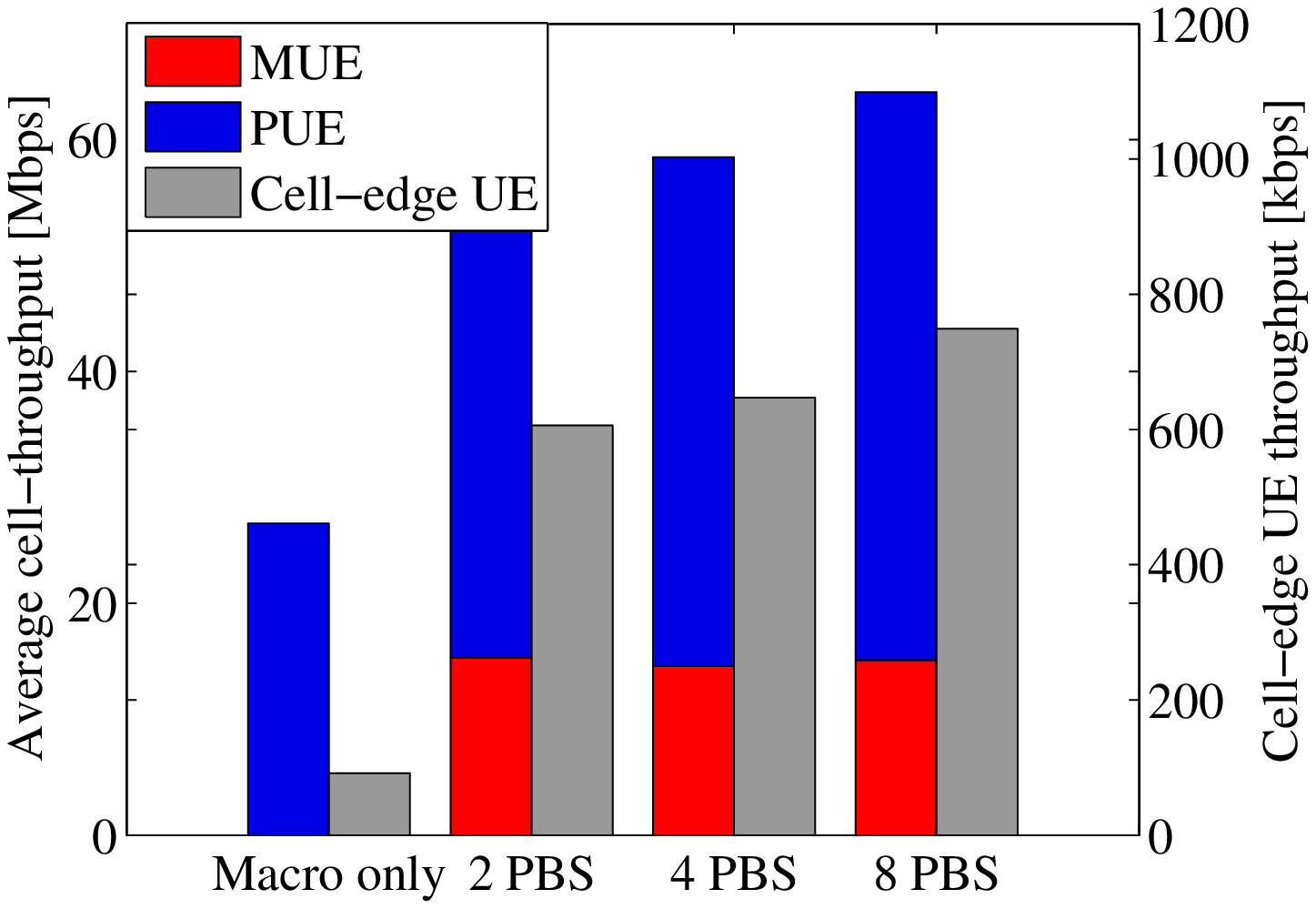}
  \put(-155,105){(a)}
\end{minipage}%
\begin{minipage}{.29\textwidth}
  \centering
  \includegraphics[width=1.0\linewidth]{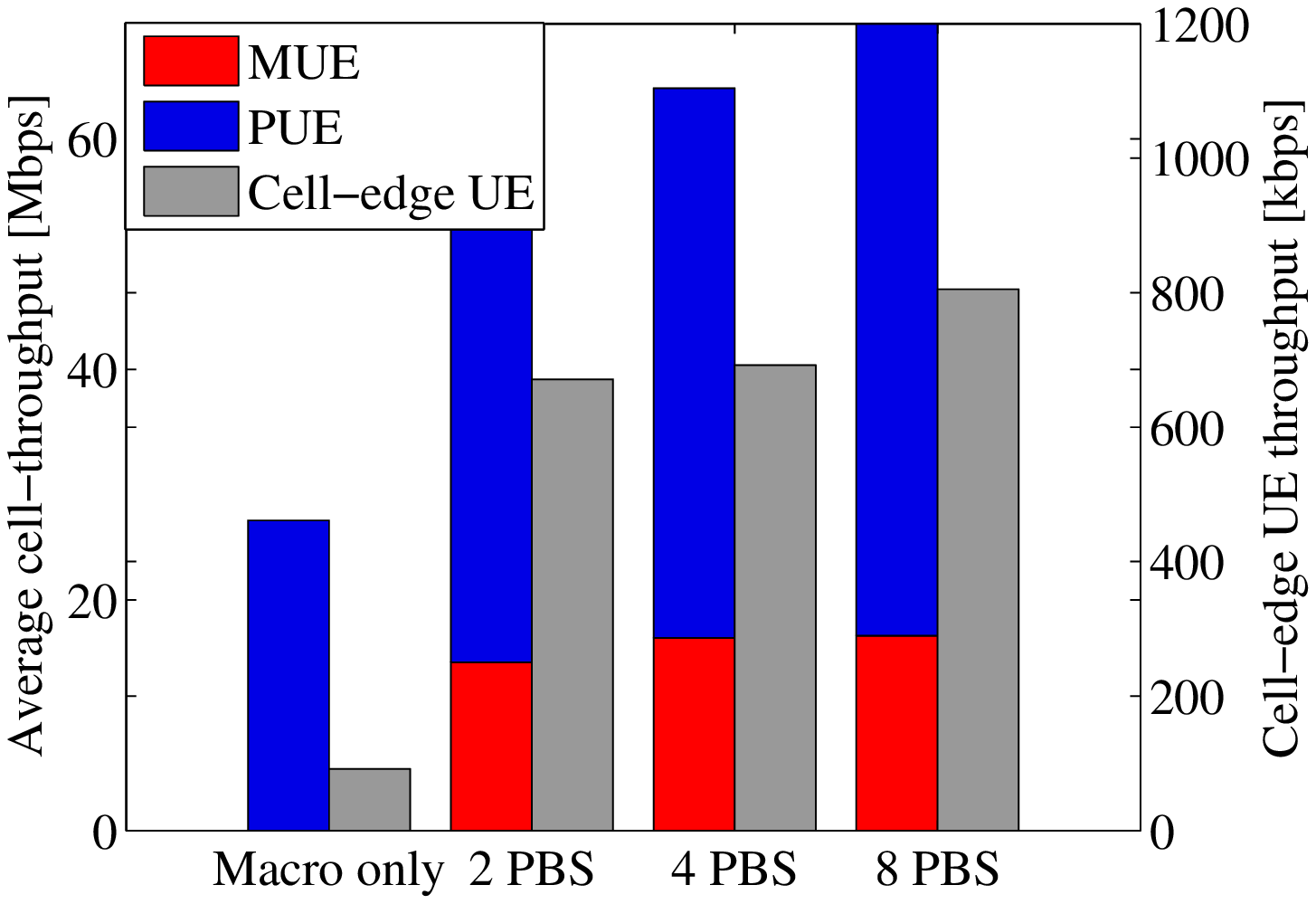}
 \put(-155,105){(b)}
\end{minipage}
\begin{minipage}{.29\textwidth}
  \centering
  \includegraphics[width=1.0\linewidth]{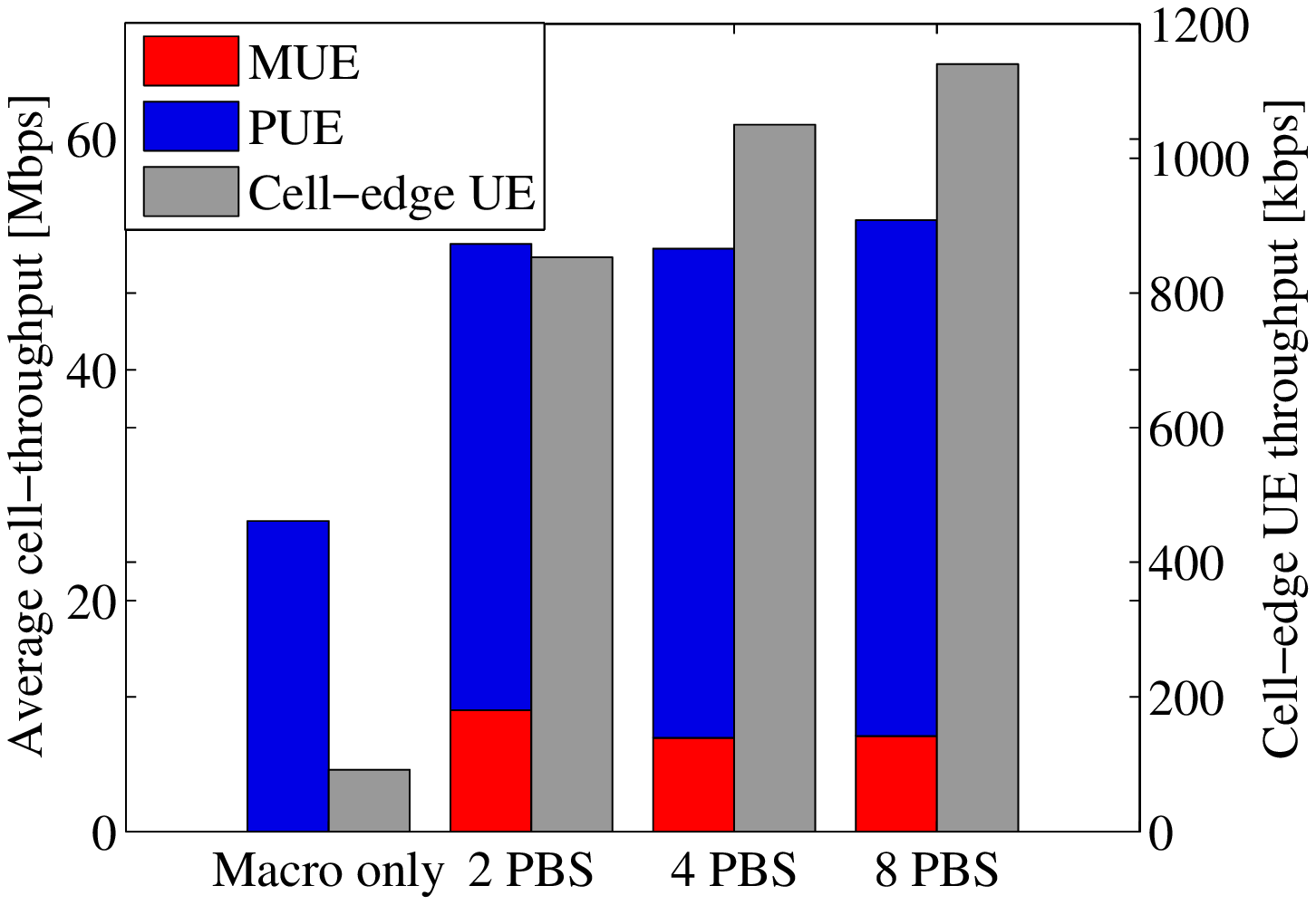}
  \put(-155,105){(c)}
\end{minipage}%
\caption{Average and cell-edge throughput versus the number of PBS per macrocell in (a) static QL based ICIC, (b) dynamic QL based ICIC, and (c) satisfactoin based ICIC.}
  \label{fig:VsPBS}
\end{figure*}

\subsection{Impact of Number of Picocells}
The impact of the number of picocells per macro sector is evaluated next.
We consider 2, 4 and 8 picocells per macro sector and compare our results with the case when no picocell is activated. Figs. \ref{fig:VsPBS} (a) - (c) 
show the results for the \emph{static QL},  \emph{dynamic QL} and  \emph{satisfaction} based algorithms, respectively. While we distinguish between picocell and macrocell average cell throughput on the left y-axis, we depict the cell-edge (5-th\%) UE throughput of all UEs in the system on the right y-axis. Hereby, the average cell throughput is the throughput per cell. From Figs. \ref{fig:VsPBS} (a) - (c) it can be observed that deploying picocells yields both average cell throughput and cell-edge UE throughput enhancement for all ICIC techniques. While the \emph{static QL} algorithm shows an approximately linearly performance increase for average cell and cell-edge UE throughput, the \emph{dynamic QL} algorithm's performance abruptly increases in the case of eight picocells per macro sector. This means a $3$-fold increase in  average cell-throughput by activating eight picocells and using the \emph{dynamic QL} ICIC technique.

The \emph{satisfaction} based ICIC technique shows the lowest dependency on the number of picocells. While the average cell throughput is slightly increased by increasing number of picocells, the cell-edge UE throughput shows an approximately linear increasing behavior. The reason is that the \emph{satisfaction} based ICIC technique does not change its strategy as long as the QoS requirements are satisfied. Comparing the cell-edge UE throughput with the $Q$-learning based ICIC algorithms, the \emph{satisfaction} based ICIC scheme provides the highest cell-edge UE throughput. This is the tradeoff of the proposed learning based ICIC schemes, in which the \emph{satisfaction} based algorithms cannot achieve very high overall performance.

\begin{figure}
	\centering
		\includegraphics[width=0.45\textwidth]{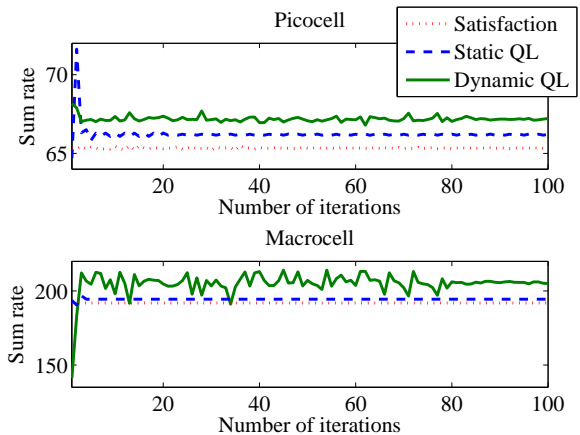}
	\caption{Convergence of the learning based time domain ICIC techniques.}
	\label{fig:convergence5}
\end{figure}

\subsection{Convergence Behavior of Time-Domain ICIC}\label{sec:Convergence}
To summarize the trade-offs of our proposed learning based time domain ICIC schemes, we show the convergence behavior of these algorithms in Fig. \ref{fig:convergence5}. All algorithms converge within a small number of iterations. While the satisfaction based approach aims at guaranteeing QoS requirements, the $Q$-learning based approaches maximize the system performance. Here, in the static $Q$-learning based approach since the macro is assumed to be static, the picocell adapts its actions very fast, which results in a fast convergence. However, since the MBS mutes on the RBs that are allocated to ER PUEs by the PBS, the macrocell performance is weak. If the MBS also performs learning in our dynamic approach more information exchange among layers is necessary that relies on high capacity backhauls and low delays. In this case the macrocell performance can be increased because of picocell offloading and optimal power allocation by both macro- and picocells. This also shows that the dynamic approach yields the best performance after convergence. The satisfaction based ICIC approach shows a fast convergence as expected, since the learning strategy will not be changed by the picocells as long as the satisfaction in terms of QoS is achieved. This has the drawback that the sum-rate cannot be maximized.

\begin{figure*}[htb!]
\centering
\begin{minipage}{.5\textwidth}
  \centering
  \includegraphics[width=1.0\linewidth]{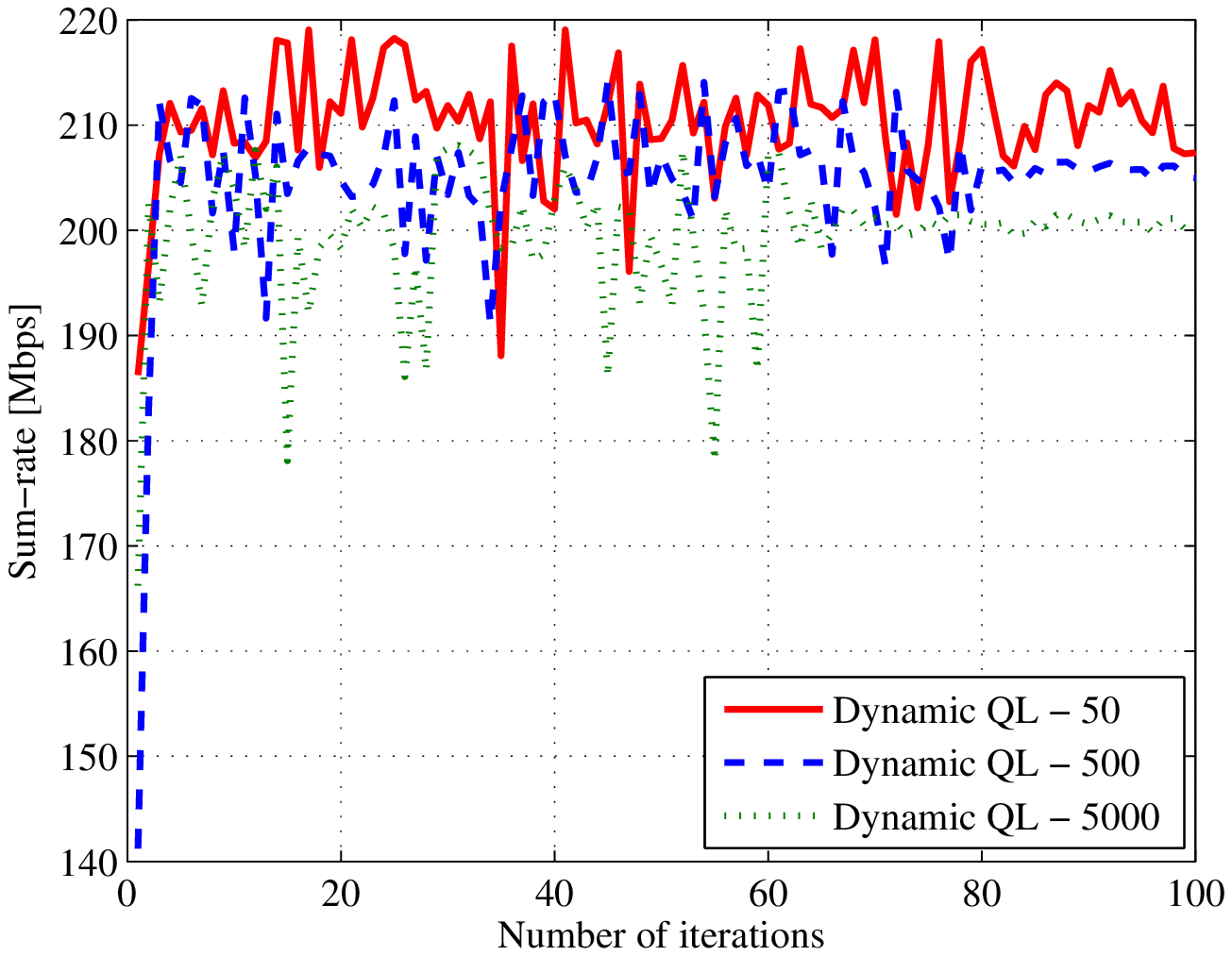}
  \put(-250,180){(a)}
\end{minipage}%
\begin{minipage}{.5\textwidth}
  \centering
  \includegraphics[width=1.0\linewidth]{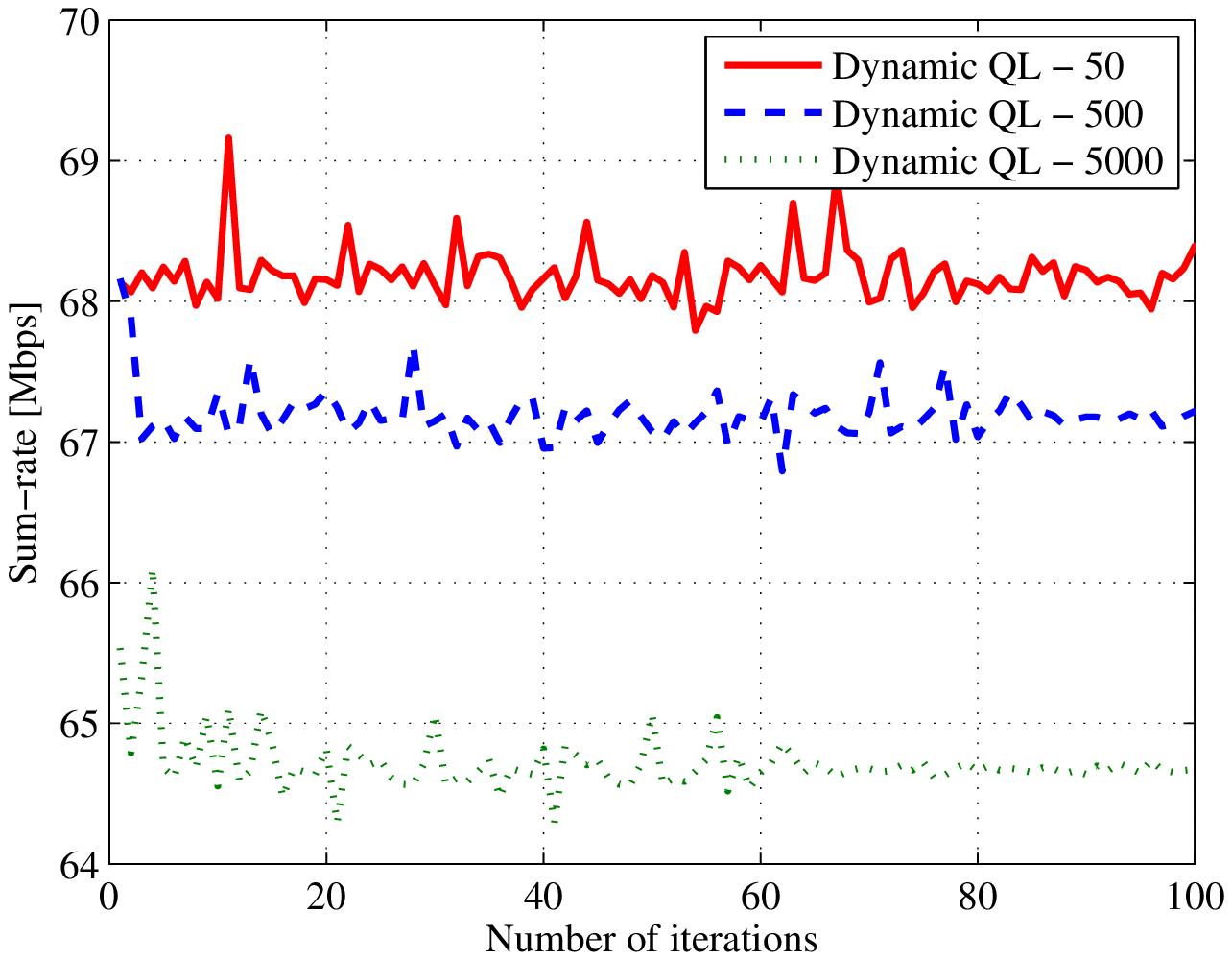}
 \put(-250,180){(b)}
\end{minipage}
\caption{Convergence of the sum-rate of dynamic QL for different cost values for (a) macrocells, and (b) picocells.}
\label{fig:ConvergenceDynQL}
\end{figure*}

In Fig.\ref{fig:ConvergenceDynQL} (a) and (b), we depict the convergence behavior of dynamic QL for different cost values in \eqref{cost}. In case of a cost value of 50, the dynamic QL approach converges slower as compared to the cost value 500 and cost value 5000, respectively. It converges to a better sum-rate than the cost value 500, but needs more iterations to converge. The simulations for a cost value of 5000 converge faster, but show a significant performance degradation.   

\begin{figure*}[htb!]
\centering
\begin{minipage}{.5\textwidth}
  \centering
  \includegraphics[width=1.0\linewidth]{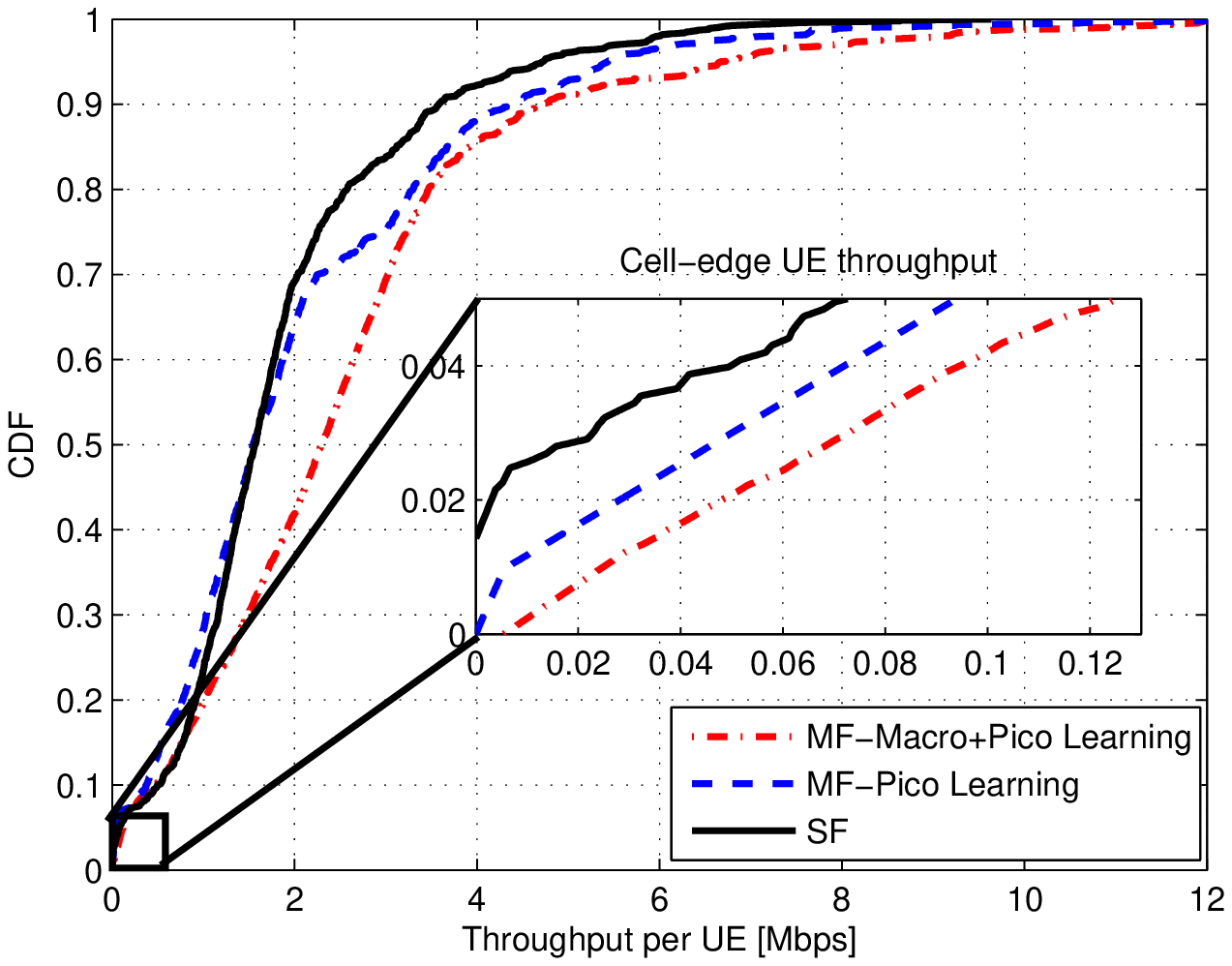}
 \captionof{figure}{CDF of the UE throughput for \\the single-flow and multi-flow ICIC learning \\algorithm for $2$ picocells per macrocell.}
  \label{fig:CDFfreqDomain}
\end{minipage}%
\begin{minipage}{.5\textwidth}
  \centering
  \includegraphics[width=1.0\linewidth]{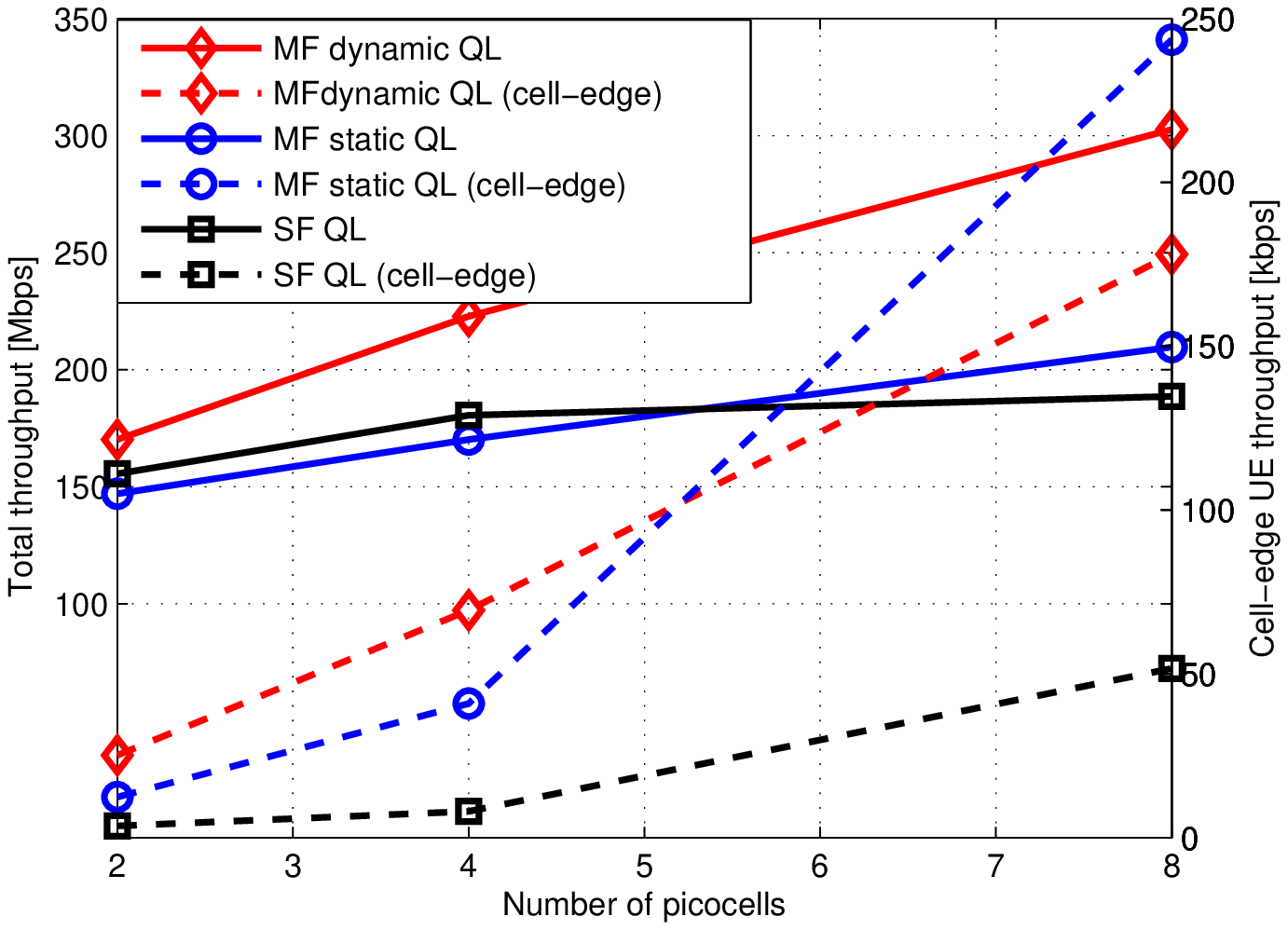}
  \captionof{figure}{Total-throughput and cell edge throughput versus the number of picocells in frequency domain ICIC.}
  \label{fig:FreqDomainVsPBS}
\end{minipage}
\end{figure*}

\subsection{Frequency Domain ICIC}
For the proposed frequency domain ICIC algorithms an analysis of the tradeoffs for single-flow CA (\emph{SF QL}) and  multi-flow CA (\emph{MF static QL} and \emph{MF dynamic QL}) is performed. Fig. \ref{fig:CDFfreqDomain} plots the UE throughput for two active picocells per macrocells. While the \emph{SF QL} and \emph{MF static QL} algorithms are in average very close to each other, the \emph{MF dynamic QL} algorithm shows a performance improvement of $47\%$ on  average. A close-up view of the cell-edge UE throughput shows that the multi-flow CA algorithms outperform the single-flow case. This is because in multi-flow CA, cell-edge UEs are served by macro- and picocell at the same time.

The behavior of the learning based frequency domain ICIC algorithms when increasing the number of picocells per macrocell is depicted in Fig. \ref{fig:FreqDomainVsPBS}. Here, the solid curves belong to the left ordinate showing the total throughput and the dashed curves refer to the right ordinate reflecting the cell-edge UE throughput. It can be observed that the \emph{MF dynamic QL} algorithm outperforms the other algorithms in terms of total throughput while the \emph{SF QL} algorithm is slightly better than the \emph{MF static QL} algorithm for less number of picocells (and vice versa for large numbers). The \emph{SF QL} algorithm shows the lowest performance for cell-edge UE throughput. It can be concluded that cell-edge UEs benefit more from multi-flow CA than from single-flow CA. Interestingly, it can be observed that the \emph{MF static QL} algorithm outperforms the \emph{MF dynamic QL} for larger number of picocells. This is because in the two-player case, the MBS cannot fully adapt to the ICIC strategies of all PBSs in the system, when the number of PBS large. 

\vspace{-0.3cm}
\section{Conclusion}\label{sec:Conclusion}
In this paper, we investigated the  performance of two-tier HetNets in which decentralized $Q$-learning and satisfaction based procedures were proposed for both time and frequency domain ICIC. The proposed approach in which PBSs optimally learn their optimal CRE bias and transmit power allocation, is shown to outperform the static ICIC solutions in time domain. While the satisfaction based approach improves the 5\% UE throughput and guarantees QoS requirements, the dynamic $Q$-learning based approach increases network capacity  relying on high capacity backhauls. In the frequency domain case, the single and multi-flow CA demonstrate that the dynamic $Q$-learning based multi-flow approach outperforms the single-flow case. Improvements of $60\%$ in the total throughput and $240\%$ in the cell-edge UE throughput are obtained in the  case of multi-flow dynamic $Q$-learning with 8 picocells per macrocell. The proposed algorithms can be extended to an $K$-tier HetNet. However, in this case it has to be defined which player selects first its primary CC and how coordination is performed. In our future work, we will extend the current framework to the non-ideal backhaul considering delays.
\vspace{-0.2cm}

\section{Appendix}
\subsection{Proof of Proposition - 1} \label{proof}
Before presenting the proof of convergence to one of the equilibrium of the game, we define the following hypothesis of the game $\mathcal G = \{\mathcal P, \{\mathcal A_p\}_{p \in \mathcal P}, \{u_p\}_{p \in \mathcal P}\}$:
\begin{enumerate}
\item The game $\mathcal G = \{\mathcal P, \{\mathcal A_p\}_{p \in \mathcal P}, \{u_p\}_{p \in \mathcal P}\}$ has at least one equilibrium in pure strategies.
\item For all $p\in \mathcal P$, it holds that $\forall a_{p'\neq p}\in \mathcal A_{p'\neq p}$, the set $u_p(p'\neq p)$ is not empty.
\item The sets $\mathcal P$ and $\{\mathcal A_p\}_{p \in \mathcal P}$ are finite.
\end{enumerate}
The first hypothesis ensures that the learning problem is well-posed, in which the players are assigned a feasible task. The second hypothesis refers to the fact that, each player is always able to find a transmit configuration with which it can be considered satisfied, given the transmit configuration of all the other players. The third hypothesis is considered in order to ensure that our algorithm is able to converge in finite time.

The proof of the proposition in section \ref{sec:satisfaction} follows from the fact that the condition $\pi_{r,n_p}^p(t_k)>0$  implies that every action profile will be played at least once
with nonzero probability during a large time interval. Because of the assumption that at least on SE exists, this action profile will be played at least once. From equation \eqref{eq:ActionSelection}, it follows that once an equilibrium is played, no player changes its current action. Thus, convergence is observed.
\subsection{Memory and Computational Requirements}\label{Complexity}

We present in what follows the memory and computational requirements of the proposed $Q$-learning and satisfaction based learning approaches when considering digital signal processors (DSPs). A theoretical estimation of the operational requirements for the mathematical operations required in the learning approaches is presented, assuming that every basic DSP instruction takes one DSP cycle~\cite{AnaThesis}.

The memory requirements of learning methods are directly related to the knowledge representation mechanisms of agents. In $Q$-learning, the agent's knowledge is represented by $Q$-tables which have the size of $|\mathcal{S}|\times |\mathcal{A}|$. In the presented two-player game this results in a memory requirement of $\left(|\mathcal{S}^m|\times |\mathcal{A}^m|+|\mathcal{S}^p|\times |\mathcal{A}^p\right)\cdot R$ memory units per game and over all RBs. In case of satisfaction based learning, the agent's knowledge is represented by the probability distribution over all actions. This results in a memory requirement of $\left(1\times |\mathcal{A}^p|\right)\cdot R$ memory units. Hence, satisfaction based learning requires significantly less memory than $Q$-learning.

\begin{figure}
	\centering
		\includegraphics[width=0.5\textwidth]{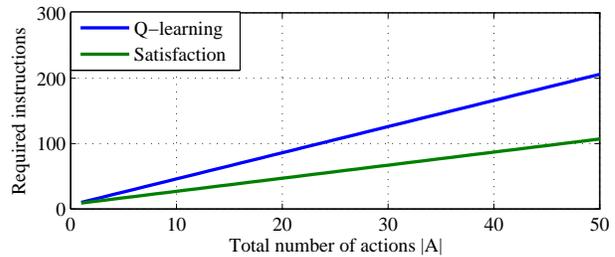}
	\caption{Required instructions of learning approaches for different number of actions.}
	\label{fig:ComputationalReq}
\end{figure}

The presented computational analysis does not take into account the compiler optimizations and the ability of DSPs to execute various instructions per clock cycle. Therefore, the analysis provides an upper bound for the computational resources that are needed by the algorithms~\cite{AnaThesis}. The computational requirements of the learning methods are given by the operations they have to execute in order to fulfill the representation of the acquired knowledge in one learning iteration. Table \ref{tab:QcompCost} summarizes the total number of operations required per RB for one $Q$-learning iteration through equation \eqref{eq:Qupdate}. 
\begin{table}
\begin{center}
\caption{Computational requirement for $Q$-learning and satisfaction based learning.}
\label{tab:QcompCost}
\begin{tabular}{|l|l|l|}\hline
\multirow{2}{*}{\textbf{Operations}} & \multicolumn{2}{c|}{\textbf{Required instructions for}}\\\hhline{~--}
&\textbf{$Q$-learning} & \textbf{satisfaction based learning}\\\hline
Identification of current and & 2 & -\\
next state in the Q-table & & \\
Memory access & $ 2\cdot |\mathcal{A}|$ & $2\cdot |\mathcal{A}|$\\
Comparison & $ 2\cdot \left(|\mathcal{A}|-1\right)$ & 2\\
Sum & 3 & 2\\
Multiplication & 2 & 2\\
Storage & 1 & 1\\
Total number of operations & $4|\mathcal{A}|+6$ & $2|\mathcal{A}|+7$\\\hline
\end{tabular}
\end{center}
\end{table}
In satisfaction based learning, one learning iteration is based on the probability update function in equation \eqref{eq:ProbUdpate}, which is only updated if the system is not satisfied. The third column of Table \ref{tab:QcompCost} summarizes for this case the total number of operations required per RB. 
The total number of operations required for $Q$-learning and satisfaction based learning is $4|\mathcal{A}|+6$ and $2|\mathcal{A}|+7$, respectively. Since $|\mathcal{A}|> 0$, satisfaction based learning requires less operations than $Q$-learning.
Fig. \ref{fig:ComputationalReq} depicts the required instructions over different number of actions for both learning approaches.

\section{Acknowledgment}
This work is supported by the SHARING project under the Finland grant 128010 and was supported in part by the U.S. National Science Foundation under the grant CNS-1406968.


\end{document}